\documentclass[trackchanges,twocolumn]{aastex701}
\usepackage{graphicx}	
\usepackage{color}
\usepackage{amsmath}	% Advanced maths commands
\usepackage{amssymb}	% Extra maths symbols
\usepackage{mathrsfs}
\usepackage{mathrsfs}
\usepackage{multirow}

\newcommand{\msun}{M_\odot}

\newcommand{\zsun}{Z_\odot}
\newcommand{\lsun}{L_\odot}

\newcommand{\cc}{{\rm cm}^{-3}}

\newcommand{\msunyr}{M_\odot~{\rm yr}^{-1}}

\newcommand{\kpc}{{\rm kpc}}
\newcommand{\mpc}{{\rm Mpc}}

\newcommand{\pc}{{\rm pc}}
\newcommand{\mum}{\mu {\rm m}}
\newcommand{\kms}{{\rm km~s}^{-1}}

\newcommand{\K}{{\rm K}}
\newcommand{\beq}{\begin{equation}}
\newcommand{\eeq}{\end{equation}}

\begin{document}

\title{Spectral Uniformity of Little Red Dots: \\A Natural Outcome of Coevolving Seed Black Holes and Nascent Starbursts}

\author[orcid=0000-0001-9840-4959,sname='']{Kohei Inayoshi}
\affiliation{Kavli Institute for Astronomy and Astrophysics, Peking University, Beijing 100871, China}
\email[show]{inayoshi@pku.edu.cn}  

\author[orcid=0000-0002-5358-5642,sname='']{Kohta Murase}
\affiliation{Department of Physics; Department of Astronomy \& Astrophysics; Center for
Multimessenger Astrophysics, Institute for Gravitation and the Cosmos, The Pennsylvania State University, University Park, PA 16802, USA}
\affiliation{Center for Gravitational Physics and Quantum Information, Yukawa Institute for Theoretical Physics, Kyoto University, Kyoto, Kyoto 606-8502, Japan}
\email[]{}

\author[orcid=0000-0003-4299-8799,sname='']{Kazumi Kashiyama}
\affiliation{Astronomical Institute, Tohoku University, Sendai, Miyagi 980-8578, Japan}
\affiliation{Kavli Institute for the Physics and Mathematics of the Universe (Kavli IPMU, WPI), The University of Tokyo, Chiba 277-8583, Japan}
\email[]{}

\begin{abstract}
The birth of seeds of massive black holes (BHs) and nascent galaxies at cosmic dawn takes place in dense gaseous environments, 
which play a crucial role in shaping their coevolution and radiation spectra.
We investigate gas accretion during the assembly of massive halos 
with $M_{\rm h}\gtrsim 10^{10-11}~\msun$ at redshifts $z\simeq 4-10$, 
driving both rapid BH feeding and concurrent nuclear starbursts.
As the BH grows to $\sim 10^{6-7}~\msun$ via super-Eddington accretion, the accretion power inflates a dense envelope 
whose effective temperature approaches the Hayashi limit at $T_{\rm eff}\simeq 5000~\K$, producing red optical emission, 
while a coeval young stellar population of $\sim 10^7~\msun$ provides blue UV emission.
This early coevolving system naturally reproduces the characteristic 
spectral features of the so-called little red dots (LRDs), a population of broad-line active galactic nuclei (AGNs), including the V-shaped UV-to-optical spectra and weakness of X-ray, infrared, and radio emission.
Massive stars in the nuclear starburst soon explode as supernovae, injecting energy and momentum that expel gas from the nucleus, quench gas supply 
to the BH envelope, and ultimately drive a transition into normal AGN phases.
For individual LRDs, the optical-to-UV luminosity ratio remains nearly constant at $L_{\rm opt}/L_{\rm UV}\simeq 2-10$ from the onset of accretion bursts for 
$\simeq 15~{\rm Myr}$, one-third of the Salpeter time, until quenching by stellar feedback.
While this ratio is sustained for the LRD population at $z\simeq 4-8$, it declines toward lower redshifts as BHs can no longer maintain red envelopes, thereby losing the LRD characteristics.
\end{abstract}


\section{Introduction}
Little Red Dots (LRDs) are a new identified class of active galactic nuclei (AGNs) uncovered by deep survey programs led with the James Webb Space Telescope (JWST) \citep[e.g.,][]{Matthee_2024,Greene_2024,Furtak_2024,Labbe_2025,Kocevski_2025},
including the highest-redshift spectroscopically confirmed LRD at $z_{\rm spec}=9.288$ \citep{Taylor_2025b} and a photometrically selected candidate at $z_{\rm phot}=10.5$ \citep{Tanaka_2025_z10LRD}.
Their properties are characterized by distinctive spectral energy distribution (SED), broad Balmer emission lines, and compact morphologies \citep[e.g.,][]{Wang_2024b,Chen_2025b,Setton_2025b,Hviding_2025}, while several canonical AGN signatures such as hard X-rays from hot coronae, near- to mid-infrared radiation from hot dust, and radio jets are weak or absent \citep[e.g.,][]{Maiolino_2025,Yue_2024,Perez-Gonzalez_2024,Akins_2025,Gloudemans_2025,Xiao_2025}.
From the observed rest-frame UV-optical-infrared fluxes, LRDs typically have luminosities of $L\sim 10^{43}-10^{44}~{\rm erg~s}^{-1}$
(uncorrected for dust extinction), which can be powered by accreting massive black holes (BHs) with masses of $M_{\rm BH}\sim 10^{6}-10^{7}~\msun$ 
or potentially even lower masses closer to the predicted mass range for seed BH populations
\citep[][references therein]{Inayoshi_ARAA_2020,Volonteri_2021}.

Their spectral uniqueness likely originates from the dense environments where the central BHs reside.
LRD spectra often show prominent absorption features superposed on broad Balmer emission lines 
\citep{Kocevski_2023,Matthee_2024,Maiolino_2024_JADES,Lin_2024,Taylor_2025a,Kocevski_2025,DEugenio_2026} together with a spectral break near the Balmer limit wavelength \citep[e.g.,][]{Greene_2024,Furtak_2024,Wang_2024b}.
These features require dense gas absorbers with high covering fractions surrounding the BH \citep{Inayoshi_Maiolino_2025}, especially for some LRDs
with an extremely deep Balmer break that is hardly explained by stellar continuum of evolved populations \citep{deGraaff_2025b,Naidu_2025,Taylor_2025b}.
High gas densities of $n_{\rm H} \gtrsim 10^{9}-10^{11}~\cc$ can collisionally excite atomic hydrogen to the $n = 2$ level 
\citep{Juodzbalis_2024}, thereby producing both Balmer absorption and break features on the transmitted AGN spectra \citep{Inayoshi_Maiolino_2025,Ji_2025}.
By analogy with stellar atmospheres, the red optical continuum can be interpreted as the Wien tail of 
a blackbody spectrum with $T_{\rm eff} \simeq 5000~\K$ without relying on dust reddening \citep{Inayoshi_2025b,Kido_2025,Liu_2025b,Begelman_Dexter_2026}.
This interpretation is supported by JWST/MIRI observations, which place stringent limits on hot dust emission in 
the rest-frame near–to-mid infrared \citep{Casey_2024,Casey_2025,Z.Li_2025,K.Chen_2025,Setton_2025b};
in some cases, warm dust emission is required to explain the infrared spectra \citep{Lyu_2024,Perez-Gonzalez_2024,Delvecchio_2025,Ronayne_2025}.
Furthermore, if the accreting BH is embedded in a dense, quasi-spherical envelope, the absence of X-rays and radio jets 
arises naturally from strong absorption and jet confinement in such environments \citep{Kido_2025}.

Independently, the BH-envelope model has been proposed in the context of seed BH formation in the early universe, 
as a pathway to the progenitors of supermassive black holes (SMBHs) that later power luminous quasars \citep[e.g.,][]{Fan_2023}.
Under circumstances with rapid gas collapse into the protogalactic nuclei, accreting BHs are embedded in dense inflows 
at accretion rates far exceeding the Eddington limit \citep{Volonteri_Rees_2005,Begelman_2006,Shang_2010,Inayoshi_Haiman_Ostriker_2016,Chon_2018,Hu_2022a,Shi_2024}.
These inflows naturally form extended envelopes with photospheric temperatures of $T_{\rm eff} \simeq 4000-7000~\K$ \citep{Begelman_2010,Hosokawa_2013,Woods_2021,Begelman_Dexter_2026}.
Such temperatures result from hydrogen recombination physics and H$^-$ opacity, similar to the surface layers of 
protostars and post-main-sequence stars \citep{Hayashi_1961}.

This dense envelope framework provides a self-consistent explanation for several key observed properties of LRDs.
However, it faces two major challenges; the origin of the rest-frame UV continuum and the presence of broad-line regions.
A cold atmosphere with $T_{\rm eff}\simeq 5000~\K$ produces neither ionizing photons nor significant UV continuum.
If broad-line emitting clouds are located deep inside the envelope, their emission is likely to be smeared out 
by multiple scattering and absorption.
Although additional UV sources such as a young stellar population can reproduce the observed UV spectra ,
the UV-to-optical flux ratio in such a simple two-component model is not naturally fixed \citep{Wang_2024b,Kocevski_2025,Ma_2025c}.
Therefore, a physical mechanism is required to maintain the observed uniformity of the LRD ``V-shaped" SED, 
ensuring that the turnover consistently occurs near the Balmer limit \citep{Setton_2025a} and the UV-to-optical flux ratio remains in a relatively narrow range.

In this paper, we investigate the properties of mass accretion in early protogalaxies with dark matter halo masses of $M_{\rm h}\gtrsim 10^{10}-10^{11}~\msun$, driving both rapid BH feeding through an inflated gaseous envelope and concurrent nuclear starbursts, and propose a two-component model to explain the V-shaped SEDs of LRD populations.
To reproduce the observed UV continuum, intense starburst activity with star formation rates (SFRs) of $\gtrsim 1~\msunyr$ is required, 
leading to the formation of a $\sim 10^7~\msun$ stellar cluster within $\sim 10$ Myr at $\lesssim 100~\pc$.
Massive stars formed in such bursts explode as supernovae and inject energy and momentum into the surrounding medium.
This feedback process expels gas from the nucleus, quenches accretion onto the BH, and thus regulates LRD activity,
leading to a transition into normal AGN phases in the subsequent accretion stages (\citealt{Inayoshi_2025a}; see also \citealt{Fujimoto_2022,Billand_2025}).
We place this scenario in the context of hierarchical structure formation and explore the key implications for LRDs, focusing on the origin of SED uniformity, 
its evolution with redshift, and the connection to parent halo properties.

%%%%%%%%%
%%% Fig. 1 %%%
%%%%%%%%%
\begin{figure*}
\centering
\includegraphics[width=155mm]{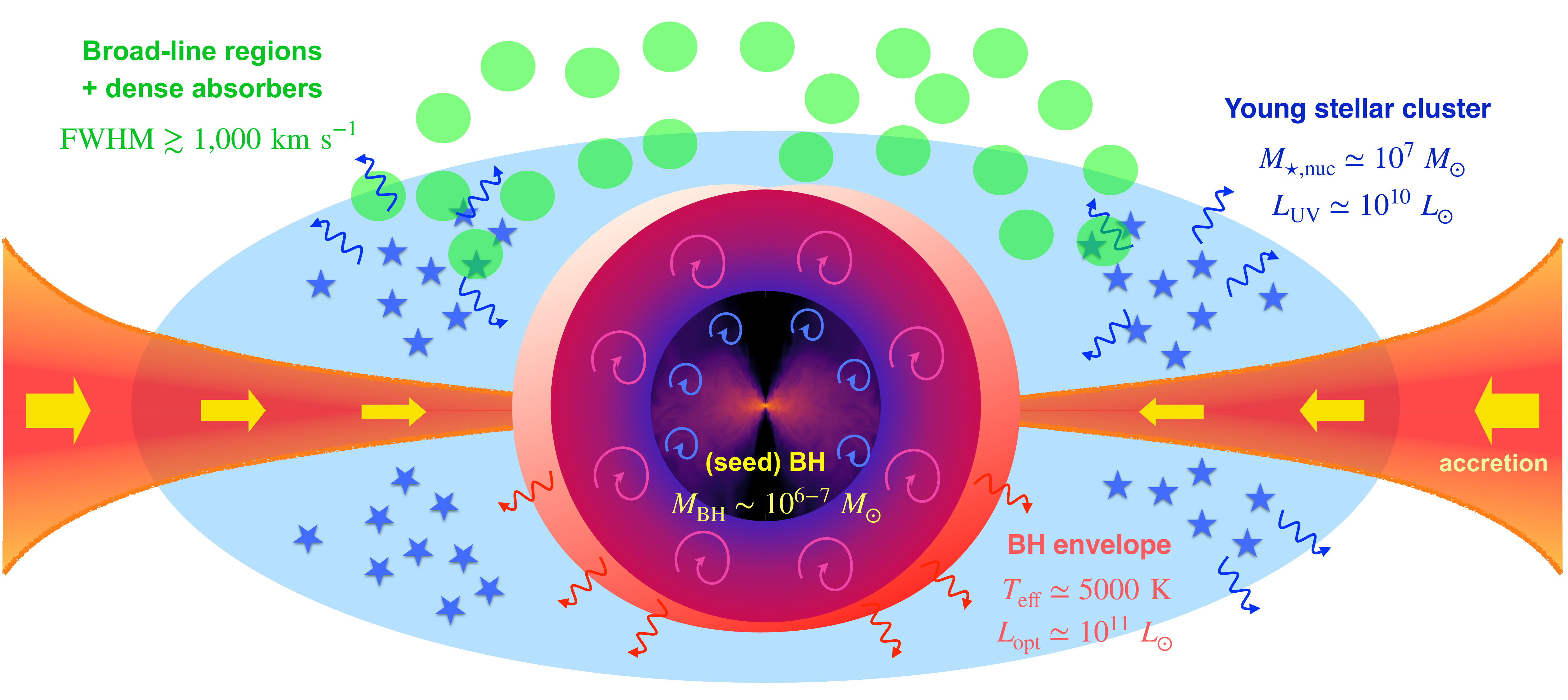}
\caption{A schematic view of the structure of an LRD powered by 
a rapidly accreting (seed) BH and young starbursts in the nuclear region.
The bloated envelope with an effective temperature of $T_{\rm eff}\simeq 5000~\K$ surrounding the BH with $M_{\rm BH}\simeq 10^6-10^7~\msun$
emits red optical emission ($L_{\rm opt}\simeq 10^{11}~\lsun$), while the compact, young stellar population with $M_{\rm \star,nuc}\simeq 10^7~\msun$
is responsible for UV emission ($L_{\rm UV}\simeq 10^{10}~\lsun$).
Ionizing radiation from massive stars illuminates broad-line region clouds with dynamical velocities of $\simeq 1,000~\kms$.
Moderate electron scattering broadens the full-width-half-maximum of emission lines to the level observed in LRD spectra.
}
\label{fig:cartoon}
\vspace{2mm}
\end{figure*}

This paper is organized as follows.
In Section~\ref{sec:LRD_SED}, we introduce our two-component SED framework, in which a BH envelope and young nuclear starbursts contribute to 
the red optical and blue UV spectral parts, respectively (Sections~\ref{sec:BHenv} and \ref{sec:OBUV}).
In Section~\ref{sec:VSED}, we demonstrate that the composite SED model successfully explains both the stacked LRD 
spectrum and NIRSpec PRISM data of individual sources.
Section~\ref{sec:LRDend} discusses the mechanisms that terminate LRD phases: supernova feedback that halts star formation and gas inflow (Section~\ref{sec:SN}), 
and self-regulation of super-Eddington accretion by powerful momentum-driven feedback from the BH to the envelope structure (Section~\ref{sec:AGNfeedback}).
In Section~\ref{sec:Lratio}, we incorporate these prescriptions into the framework of high-$z$ halo assembly and show that feedback in the nucleus preserves the optical-to-UV luminosity ratio of LRD populations
at $z\sim 4-8$, while this ratio decreases at lower redshifts, gradually erasing the distinctive LRD colors and reducing their abundance.
In Section~\ref{sec:discussion}, we extend the model to address other LRD properties, including the origin of the broad-line region (Section~\ref{sec:BLR}),
the absence of high-energy emission lines in the observed spectra (Section~\ref{sec:ionph}), the X-ray emission (Section~\ref{sec:Xray}), the weakness of time variability (Section~\ref{sec:Tvar}), the LRD abundance (Section~\ref{sec:phi}), 
and metal-enrichment processes (Section~\ref{sec:metal}).
We summarize our findings in Section~\ref{sec:summary}.

\vspace{2mm}
\section{Spectra of Newly-born Seed BHs}\label{sec:LRD_SED}

Figure~\ref{fig:cartoon} illustrates a schematic view of the nuclear structure of an LRD discussed in this paper.
In this framework, LRD radiation is powered by a rapidly accreting (seed) BH and young starbursts in the nuclear region.
As a short summary, the bloated envelope with an effective temperature of $T_{\rm eff}\simeq 5000~\K$ surrounding the BH
emits red optical emission (Section~\ref{sec:BHenv}), while the compact, young stellar population with $\sim 10^7~\msun$ is responsible for UV emission (Section~\ref{sec:OBUV}).
The two components of radiation sources can reproduce the V-shaped SED observed in LRDs, as discussed in Section~\ref{sec:VSED}.
Ionizing radiation from massive stars illuminates broad-line region clouds with dynamical velocities of $\simeq 1,000~\kms$.
Moderate electron scattering broadens the full-width-half-maximum of emission lines to the level observed in LRD spectra
(Section~\ref{sec:BLR}).

\subsection{Rapidly accreting envelope onto a BH}\label{sec:BHenv}

Our model assumes that LRDs are powered by accreting massive BHs enshrouded by a dense, extended gaseous envelope. 
This envelope forms a bloated, cold atmosphere with an effective temperature of $T_{\rm eff} \sim 5000~\K$.
If the radiative luminosity from the envelope surface ($L_{\rm ph}$) is limited to the Eddington luminosity ($L_{\rm Edd}$) or a similar value, 
the photospheric radius can be expressed as
\begin{align}
R_{\rm ph} \simeq 1.7\times 10^{16}~{\rm cm}~\lambda_{\rm Edd}^{1/2} M_6^{1/2}T_0^{-2},
\end{align}
where $\lambda_{\rm Edd}\equiv L_{\rm ph}/L_{\rm Edd}$, $M_6=M_{\rm BH}/(10^6~\msun)$, and $T_0 =T_{\rm eff}/(5000~\K)$.
The circular velocity around the photospheric radius is 
\begin{align}
V_{\rm ph}=\sqrt{\frac{GM_{\rm BH}}{R_{\rm ph}}} 
 \simeq 3.0\times 10^{-3}c~\lambda_{\rm Edd}^{-1/4} M_6^{1/4}T_0,
 \label{eq:Vph}
\end{align}
where $G$ is the gravitational constant and $c$ is the speed of light.
Here, the envelope mass is not included as its mass is lower or at most comparable to the BH mass \citep{Kido_2025}.
The surface gravity on the envelope ($g \equiv GM_{\rm BH}/R_{\rm ph}^2$) is calculated as 
\begin{align}
g= \frac{\kappa_{\rm es}\sigma T_{\rm eff}^4}{\lambda_{\rm Edd}c}
\simeq 0.47~{\rm cm~s}^{-2}~\lambda_{\rm Edd}^{-1}T_0^4,
\end{align}
similar to the typical surface gravity on super-giant stars, where $\sigma$ is the Stefan-Boltzmann constant and $\kappa_{\rm es}$ is the electron scattering opacity.

The envelope is supported by radiation pressure powered by BH accretion.
In the nearly adiabatic regime (minimal radiative energy losses), the structure resembles convection-dominated accretion flows 
(CDAFs; \citealt{Quataert_2000}), in which inflows and outflows coexist.
Convective eddies reduce the inward mass flux toward smaller radii, following a power-law form of $\dot{M}_{\rm in}\propto r^p$, with 
$p\simeq 0.5-0.7$ observed in numerical simulations \citep{Stone_1999,Yuan_Narayan_2014,Inayoshi_2018}.
As a result, even if the large-scale mass inflow rate greatly exceeds the Eddington value, the BH feeding rate self-regulates to 
$\dot{M}_{\rm BH} \sim \mathcal{O}(\dot{M}_{\rm Edd})$ \citep[e.g.,][]{Hu_2022a}, and the emergent luminosity remains near $L_{\rm ph} \simeq L_{\rm Edd}$.
In this framework, the red optical continua of LRDs arise from the Wien tail of the blackbody spectrum emitted by the envelope surface, 
without imposing dust reddening.
The observed rest-frame optical luminosity ($L_{\rm opt} \simeq 10^{44}~{\rm ergs}^{-1}$; \citealt{Akins_2025}) can be reproduced 
by the Eddington luminosity of a BH with $M_{\rm BH} \simeq 10^6~\msun$ for $A_V = 0$.

To maintain a radiation-dominated convective envelope against disruption by AGN feedback, the mass inflow rate from 
the parent halo must far exceed the Eddington rate (see discussion in Section~\ref{sec:AGNfeedback}), defined as $\dot{M}_{\rm Edd}\equiv L_{\rm Edd}/(\eta_{\rm rad}c^2)$ with a 10\% radiative efficiency $\eta_{\rm rad}=0.1$.
Such high inflow rates are plausible in protogalactic nuclei hosted by massive dark matter halos with 
masses of $M_{\rm h}$, in which the baryonic accretion rate with a cosmic baryon fraction of $f_{\rm b} = 0.16$, can reach
\begin{equation}
\dot{M}_{\rm b}\simeq 1.5\times 10^3~\msunyr f_{\rm b,0.16}M_{\rm h,12}^{\beta+1} \left(\frac{1+z}{10}\right)^{5/2},
\label{eq:Mdotb}
\end{equation}
where $M_{\rm h,12}=M_{\rm h}/(10^{12}~\msun)$ and $f_{\rm b,0.16}=f_{\rm b}/0.16$ \citep{Dekel_2013}. 
The halo mass and redshift dependence is calibrated so that the halo mass growth becomes consistent with the Press-Schechter 
formalism and cosmological N-body simulations \citep{Wechsler_2002,Fakhouri_2010}.
While the index value is originally given as $\beta\simeq 0.14$ from fitting of halo merger trees \citep{Dekel_2013},
$\beta\simeq 0.03$ is found to provide a better approximation for halos of $\sim 10^{11}~\msun$ at $z>5$ \citep{Dekel_2025b}.
In this paper, we fix $\beta=0.03$ for all the mass range over redshift.

It is worth noting that the formula for the mass accretion rate can be further simplified if $\beta=0$ as,
\begin{equation}
M_{\rm h}(z) =M_{\rm h,0}\cdot e^{-k_{\rm h}(z-z_{\rm 0})},   
\end{equation}
where $M_{\rm h,0}$ is the halo mass at $z=z_0$ and $k_{\rm h}$ represents the growth rate of the dark matter halo.
The growth rate estimated from merger-tree calculations for massive halos of $M_{\rm h}\simeq 10^{12}~\msun$ at $z=6$ ranges over 
$0.25\lesssim k_{\rm h}\lesssim 1.15$ \citep{Hu_2025}.
Moreover, under the same approximation, the baryonic inflow rate is expressed as
\begin{equation}
\dot{M}_{\rm b}\simeq 144~\msunyr f_{\rm b,0.16}T_{\rm vir,6}^{3/2} \left(\frac{1+z}{10}\right),
\label{eq:Mdotb2}
\end{equation}
where $T_{\rm vir,6}=T_{\rm vir}/(10^6~\K)$ is the normalized halo virial temperature. 
Therefore, if the gas supply feeds the nuclear region of a massive halo with a virial temperature of $T_{\rm vir,6}\simeq 1$, this rate corresponds to $\sim \mathcal{O}(100)\times \dot{M}_{\rm Edd}$ for a BH with $M_{\rm BH} \lesssim 10^7~\msun$.
Such an inflow is sufficient to confine ionizing radiation from the central AGN through efficient radiative recombination and the strong ram pressure 
of the accreting material \citep{Inayoshi_Haiman_Ostriker_2016,Sakurai_2016,Takeo_2018}, thereby leading to the formation of a cold atmospheric envelope 
rather than the photoionized region.

The two expressions of Equations~(\ref{eq:Mdotb}) with $\beta=0.03$ and (\ref{eq:Mdotb2}) yield similar inflow rates within 10\% differences.
Such a small difference does not affect our discussion of the inflow rate itself, so we adopt the simplified form (Equation~\ref{eq:Mdotb2}) as this expression also makes the formulation of other quantities more tractable.
However, when integrating the halo mass growth, the cumulative effect of these errors becomes significant, and we therefore employ the more accurate expression (Equation~\ref{eq:Mdotb}) with $\beta=0.03$ for that purpose.

\subsection{Nuclear starbursts}\label{sec:OBUV}

%%%%%%%%%
%%% Fig. 2 %%%
%%%%%%%%%
\begin{figure*}
\centering
\includegraphics[width=84mm]{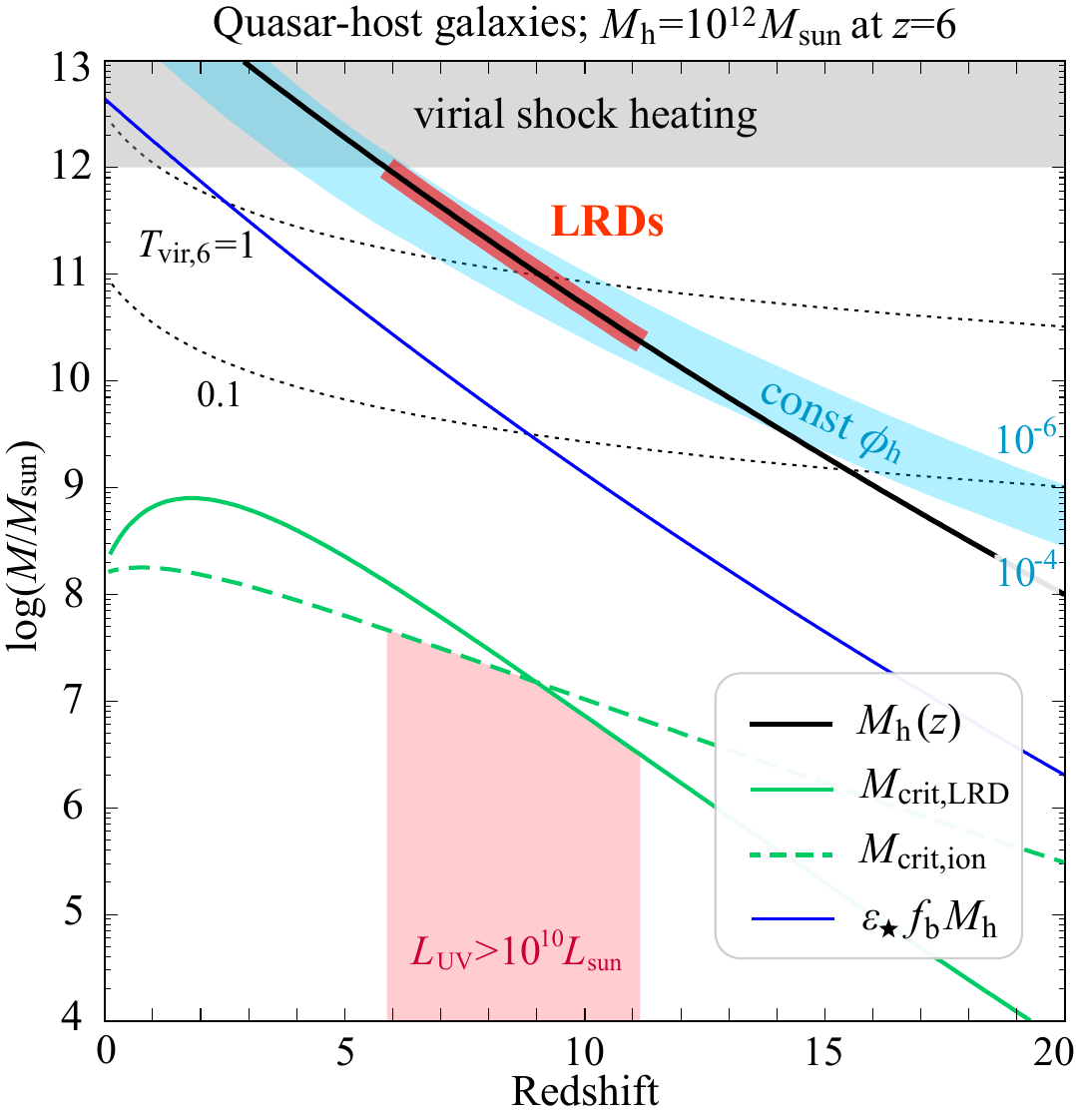}\hspace{6mm}
\includegraphics[width=84mm]{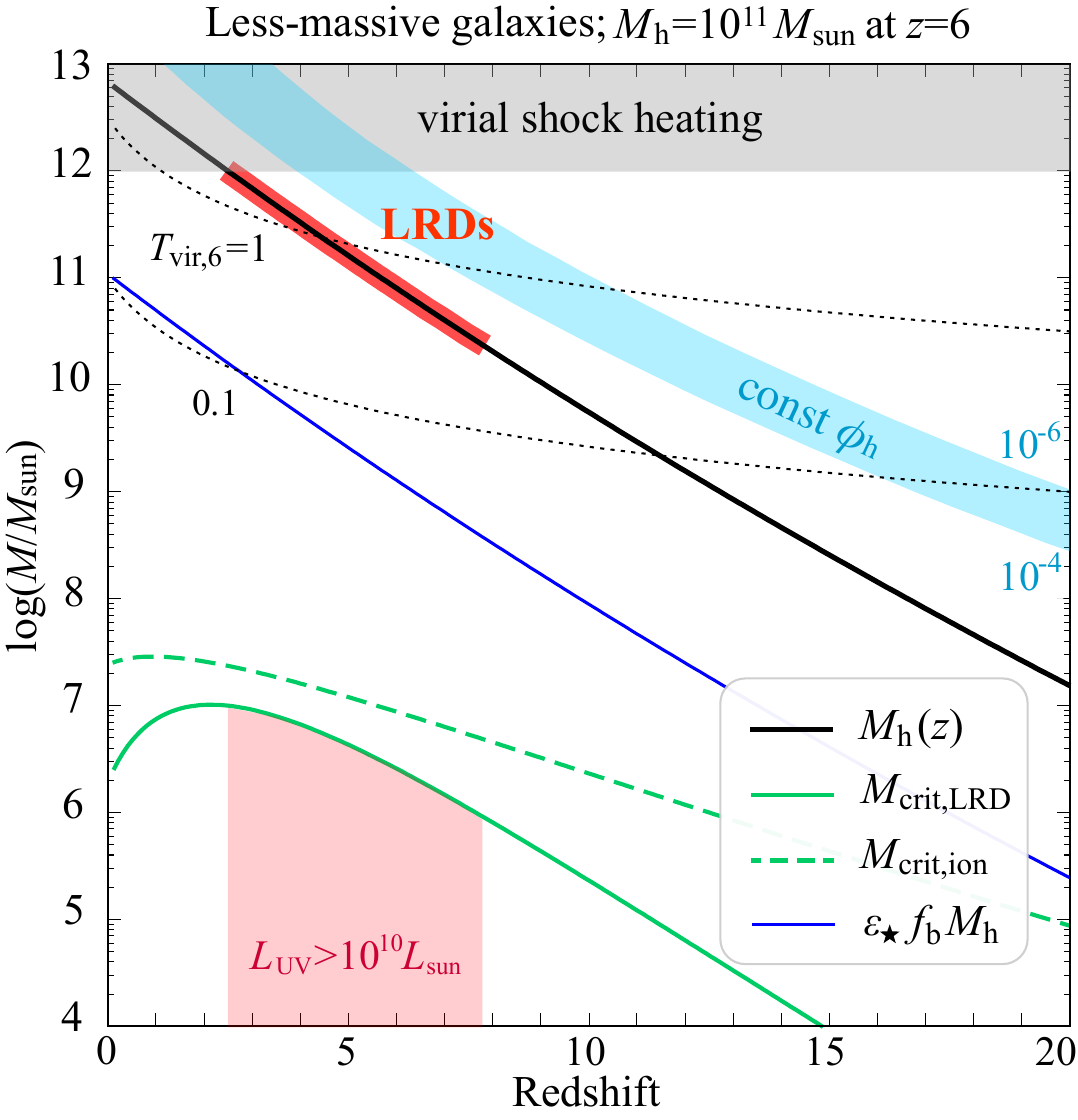}
\caption{Critical mass scales of dark matter halos, stars, and BHs as a function of redshift,
for two cases of halo-mass assembly history $M_{\rm h}(z)$ (black solid) where 
the halo mass reaches $M_{\rm h}\simeq 10^{12}$ (left) and $10^{11}~\msun$ (right) at $z=6$.
The black dotted curves present the halo masses for fixed virial temperatures at $T_{\rm vir} =10^5$ and $10^6~\K$,
and the blue shaded region indicates the range of cosmic halo abundance between $\phi_{\rm h}(\geq M_{\rm h})=10^{-6}$ and $10^{-4}~\mpc^{-3}$.
The red thick curve and shaded region indicate the conditions for active star formation (${\rm SFR}_{\rm nuc}\gtrsim 1~\msunyr$) yielding UV luminosities 
of $L_{\rm UV} \gtrsim 10^{10}~\lsun$, bright enough to explain the UV emission of LRDs.
The critical BH masses against momentum feedback ($M_{\rm crit,LRD}$) 
and radiative feedback ($\dot{M}_{\rm crit,ion}$), and 
the upper limit of stellar mass ($\epsilon_\star f_{\rm b}M_{\rm h}$) are 
shown along the halo assembly.
}
\label{fig:z_growth}
\vspace{3mm}
\end{figure*}

Next, we consider star formation episodes in the nuclear scale within massive dark matter halos with $M_{\rm h}$ at redshift $z$, and discuss the origin of blue UV emission of LRDs.
For $T_{\rm vir} \simeq 10^6~\K$, the corresponding halo mass, virial radius, and gravitational binding energy are
\begin{equation}
M_{\rm h} \simeq 9.5\times 10^{10}~\msun~T_{\rm vir,6}^{3/2}\left(\frac{1+z}{10}\right)^{-3/2},
\end{equation}
\begin{equation}
R_{\rm vir} \simeq 14.7~\kpc~T_{\rm vir,6}^{1/2}\left(\frac{1+z}{10}\right)^{-3/2},
\end{equation}
\begin{equation}
E_{\rm bin,h}  \simeq 2.6\times 10^{58}~{\rm erg} ~T_{\rm vir,6}^{5/2}\left(\frac{1+z}{10}\right)^{-3/2},
\end{equation}
where the mean molecular weight is set to $\mu=0.6$ \citep{Bryan_Norman_1998}.

In Figure~\ref{fig:z_growth}, we present the halo mass for fixed virial temperatures at $T_{\rm vir} =10^5$ and $10^6~\K$ (dotted curves).
Two cases of halo-mass assembly history, $M_{\rm h}(z)$, are shown (black solid curve) by integrating the mass growth rate 
of Equation~(\ref{eq:Mdotb}) from $z\simeq 20$, where the mass normalization is set so that the halo mass reaches 
$M_{\rm h}\simeq 10^{12}$ (left panel) and $10^{11}~\msun$ (right panel) at $z=6$, respectively.
The more massive case corresponds to the halo assembly of quasar host galaxies at $z\simeq 6$,
with a cosmic abundance of $\phi_{\rm h} \lesssim 10^{-4}~\mpc^{-3}$ (blue shaded region), while the less massive case represents more abundant 
galaxies at similar redshifts.
Our fiducial halo mass is well below the critical value for virial shock heating,
$\gtrsim 10^{12}~\msun$ (gray shaded region), where the shock heating rate exceeds the bremsstrahlung 
cooling rate in the hot plasma (\citealt{Dekel_2006}, see also \citealt{Rees_Ostriker_1977}).
Therefore, the baryonic matter within the halo of $<10^{12}~\msun$ can flow inward efficiently as cold streams \citep{Keres_2005,Dekel_2006,DiMatteo_2012}.

Under the conditions for runaway contraction, where the radiative cooling timescale is shorter than the dynamical and sound-crossing timescales
\citep{Rees_Ostriker_1977}, cold streams feed the central region of a halo until non-negligible angular momentum of gas forms
a rotationally-supported disk.
For simplicity, we follow the disk model provided by \citet{Mo_1998}, and assume that the the gas settles to an isothermal disk
with a temperature $T\simeq 10^4~\K$ and a mass fraction $f_{\rm d}$
within a halo with of $T_{\rm vir}\gg T$. 
Thus, the disk mass is expressed as
\begin{align}
    M_{\rm d}\simeq 4.6\times 10^{9}~\msun~f_{\rm d,0.3}T_{\rm vir,6}^{3/2}\left(\frac{1+z}{10}\right)^{-3/2},
\label{eq:Mdisk}
\end{align}
where the disk mass fraction is set to $f_{\rm d}=0.3f_{\rm d,0.3}$ ($f_{\rm b,0.16}$ is omitted hereafter).
The disk mass is built up through intense cold-stream accretion from the halo, on a timescale of $M_{\rm d}/\dot{M}_{\rm b}\simeq 30~{\rm Myr}~f_{\rm d,0.3}[(1+z)/10]^{-5/2}$.
The disk formation timescale is sufficiently shorter than the cosmic time interval for a significant redshift evolution ($\Delta z \simeq 1$), 
but is comparable to the key timescales for BH growth, stellar feedback, and regulation of LRD activity, 
each of which is $\sim 30-40\%$ of the Salpeter time (see Section~\ref{sec:LRDend}).
Therefore, the nuclear disk structure is in the place at the onset of the LRD phase, essentially concurrent with starbursts and BH feeding episodes.

We here assume that the gas conserves the same specific angular momentum as dark matter when forming a disk 
\citep[e.g.,][]{Fall_Romanowsky_2018,Romeo_2020,Romeo_2023}, yielding
\begin{equation}
 R_{\rm d} \equiv \lambda_{\rm dm} R_{\rm vir} \simeq 740~\pc~\bar{\lambda}T_{\rm vir,6}^{1/2}\left(\frac{1+z}{10}\right)^{-3/2},
\end{equation}
where $\lambda_{\rm dm}$ is the spin parameter of dark matter, known to follow 
a log-normal distribution function with a mean $\lambda_0=0.05$ and dispersion $\sigma_\lambda = 0.5$ \citep[e.g.,][]{Bullock_2001}, and 
$\bar{\lambda}=\lambda_{\rm dm}/\lambda_0$ is the spin parameter normalized by the mean value.
\citet{Rinaldi_2025} reported a morphological analysis of 99 LRDs, finding that 
$\sim 70\%$ appear predominantly compact $(\lesssim 400~\pc$) even when stacking the short-wavelength bands covering the rest-frame UV.
The remaining 30\% exhibit extended components; 50\% show at least two distinct associated sources or galaxy components, while the rest appear as single sources with highly asymmetric structures.
Similar complex UV morphologies have also been independently observed in \citet{Chen_2025a,Chen_2025b}, where the SEDs of off-centered components relative to 
the central point source are better explained by nebular emission powered by the central AGN, such as narrow-line regions.
These new measurements of LRD sizes, especially in the UV, suggest that star formation is triggered preferentially on nuclear scales ($R_{\rm nuc}<R_{\rm d}$), growing in an inside-out fashion.
This interpretation naturally explains why the stellar component does not appear significantly extended in current observations.
Alternatively, the nascent stellar disk is intrinsically compact, possibly reflecting a lower spin parameter of the dark matter halo, though the slowly-spinning halo abundance is lower  ($\lambda_{\rm dm}<\lambda_0$; see discussion in \citealt{Pacucci_Loeb_2025}).
While we adopt the mean spin parameter ($\lambda_{\rm dm}=\lambda_0$) throughout this paper, our model explains the fact that the LRD abundance is substantially lower than the dark matter halo abundance, owing to the $\sim 1\%$ duty cycle for the LRD phase regulated by stellar feedback
(see discussion in Section~\ref{sec:phi}).

In the early stage of disk formation, the gas density profile is assumed to be an exponential disk along the radial direction $r$ supported by 
gas pressure in the vertical $z$-direction, as 
$n(r)=n_0 \exp(-2r/R_{\rm d}){\rm sech}^2(z/\sqrt{2}z_0)$, where $R_{\rm d}$ is the disk size and $z_0$ is the scale height \citep{Oh_Haiman_2002}.
The characteristic density $n_0$ is obtained by comparison between the integrated disk mass $M_{\rm d}$ from Equation~(\ref{eq:Mdisk}),
yielding 
\begin{equation}
    n_0\simeq 1.8\times 10^3~\cc ~f_{\rm d,0.3}^2 \bar{\lambda}^{-4} \left(\frac{T_{\rm vir}}{T}\right)\left(\frac{1+z}{10}\right)^{3}.
\end{equation}
Such a gas-rich disk is prone to gravitational instability, which can trigger bar-mode angular momentum 
transport and/or disk fragmentation into clumps.
The instability is characterized in terms of the Toomre's $Q$ parameter, evaluated at $r=R_{\rm d}$ as 
\begin{equation}
    Q=\frac{c_{\rm s}\kappa}{\pi G\Sigma_{\rm d}},
\end{equation}
where $\Sigma_{\rm d}$ is the surface density of the disk, and $\kappa$ is the epicyclic frequency, 
which is related to the orbital frequency $\Omega$ as $\kappa^2 = rd\Omega^2/dr + 4\Omega^2$,
where $\Omega=V_{\rm c}(r)/r$ and $V_{\rm c}(r)$ is the circular velocity at $r$ contributed from 
the disk and dark matter \citep{Mo_1998}.
The disk instability likely occurs for low spin parameters because the disk has higher surface density and 
the Toomre-$Q$ value results in $Q\lesssim 1$.
\citet{Oh_Haiman_2002} calculated the fraction of halos where gaseous disks become gravitationally unstable
by integrating over the log-normal type spin parameter distribution function with a mean $\lambda_0=0.05$
and dispersion $\sigma_\lambda = 0.5$ \citep[e.g.,][]{Bullock_2001}.
They found that even though high-spin halos are considered ($\lambda_{\rm dm}\gg \lambda_0$), 
$\gtrsim 70\%$ (90\%) of gaseous disks embedded in halos with $T_{\rm vir}=10^{5.5}~(10^6)~\K$ lead to $Q\lesssim 1$,
indicating that fragmentation and star formation are inevitable in our cases.

%%%%%%%%%%
%%% Fig. 3 %%%
%%%%%%%%%%
\begin{figure*}
\centering
\includegraphics[width=86mm]{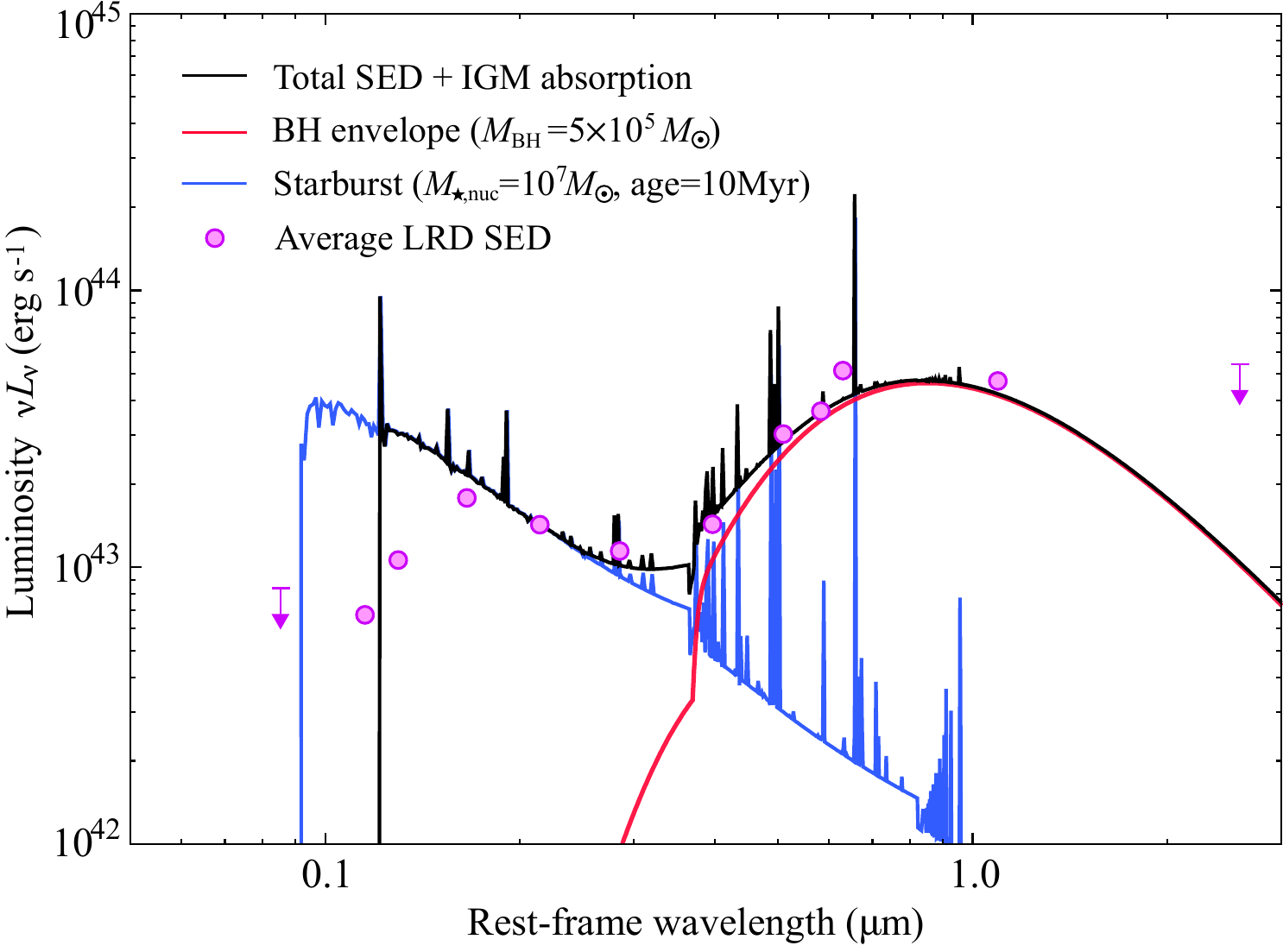}\hspace{5mm}
\includegraphics[width=87mm]{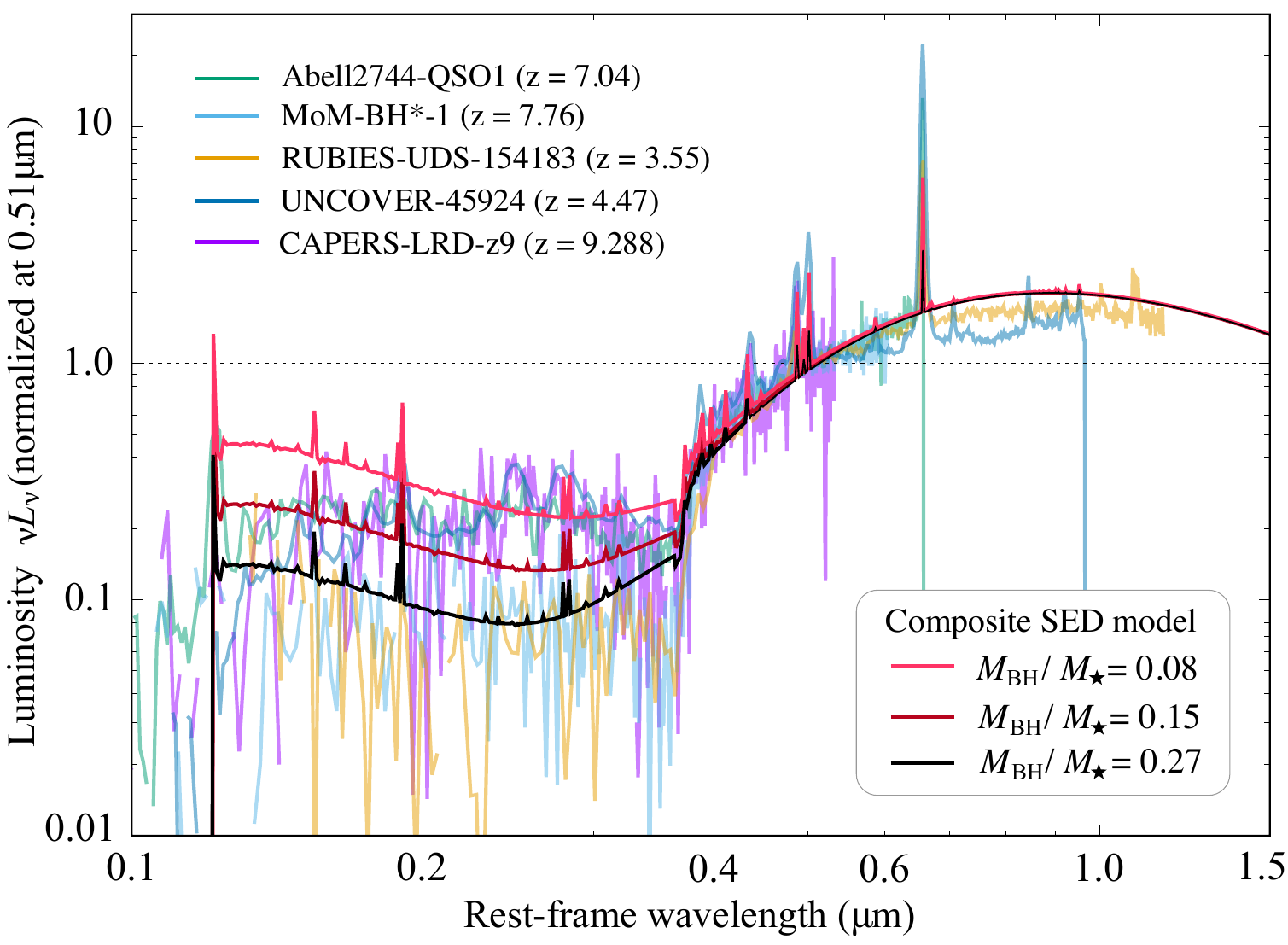}
\caption{{\it Left}: Model SED for an LRD (black), consisting of two components:
a BH envelope with $T_{\rm eff}=4300~\K$ with the Eddington luminosity of a $M_{\rm BH}=5\times 10^5~\msun$ BH (red), 
and a 10~Myr-old young starburst with ${\rm SFR}=1.0~\msunyr$ (blue).
The average photometric SED of $z\sim 6$ LRDs obtained from \citet{Akins_2025} is overlaid for comparison.
{\it Right}: Model SEDs of a LRD with a mass ratio of $M_{\rm BH}/M_{\rm \star,nuc}\simeq 0.27$, $0.15$, and $0.08$,
where the envelope luminosity is assumed to be equivalent to the BH Eddington luminosity and the stellar age is 10 Myr.
The NIRSpec MSA/PRISM spectra of several LRDs normalized at rest-frame $0.51~\mum$ are overlaid:
A2744-QSO1 at $z = 7.04$ \citep{Furtak_2024,Ji_2025}, MoM-BH$^\ast$-1 at $z = 7.76$ \citep{Naidu_2025},
RUBIES-UDS-154183 at $z = 3.55$ \citep{deGraaff_2025b}, UNCOVER-45924 at $z = 4.47$ \citep{Labbe_2024b},
and CAPERS-LRD-z9 at $z=9.288$ \citep{Taylor_2025b}.
}
\label{fig:sed_fid}
\vspace{5mm}
\end{figure*}

For simplicity, we assume a constant star formation efficiency $\epsilon_\star$ and characteristic timescale $\tau_\star$,
so that the star formation rate in the disk is given as ${\rm SFR}(<r)=\epsilon_\star M_{\rm d}(<r)/\tau_\star$.
For $\tau_\star \lesssim 10-100~{\rm Myr}$, intense starburst episodes are expected in the inner disk, and stars and their clusters formed 
in the outskirts may migrate inward via disk interactions \citep{Inayoshi_Haiman_2014,Regan_2014,Ceverino_2015,Chon_Omukai_2020,Dekel_2025a,Yajima_2025}.
In a self-gravitating, nearly isothermal gaseous disk,
the density radial distribution evolves from $n(r)\simeq n_0$ to a centrally-concentrated profile of $n(r)\propto r^{-2}$ within the disk, yielding a disk enclosed mass of $M_{\rm d}(<r) \propto r$.
In this case, the star formation rate inside $R_{\rm nuc} \equiv f_{\rm n} R_{\rm d}$ is
\begin{align}
    \frac{{\rm SFR}_{\rm nuc}}{\msunyr} 
    & \simeq 4.6 ~\zeta T_{\rm vir,6}^{3/2}\left(\frac{1+z}{10}\right)^{-3/2},
    \label{eq:SFR}
\end{align}
where $\zeta \equiv \epsilon_{-1} \tau_{1}^{-1} f_{\rm d,0.3}f_{\rm n,0.1}$, $\epsilon_{-1} = \epsilon_{\star}/0.1$, 
$\tau_{\rm  1}=\tau_{\rm \star}/10~{\rm Myr}$, and $f_{\rm n,0.1}=f_{\rm n}/0.1$.
The choice of $f_{\rm n}$ is supported by two considerations.
First, the star-forming region must be confined within the point-spread function of JWST/NIRCam, rendering LRDs spatially unresolved point sources 
with $R_{\rm nuc}\lesssim 50-100~\pc$, even in cases of strong lensing magnification \citep[e.g.,][]{Furtak_2024,Labbe_2025}.
Second, star formation proceeds in an inside-out fashion, initially concentrated in the central region rather than across the entire disk.
In nearby disk galaxies, the bulge-to-disk size ratio is typically $\lesssim 10\%$ (and up to $\sim 20\%$ in bulge-dominated systems) 
\citep[e.g.,][]{Courteau_1996,MacArthur_2003,Kormendy_Kennicutt_2004}.
This fact is consistent with the view that early the first starburst episodes primarily build the bulge component, as suggested 
by the older stellar populations typically found in bulges compared to disks.

For a given nuclear SFR, the corresponding UV luminosity is calculated as
\begin{equation}
    L_{\rm UV}    = 10^{10}~\lsun \left(\frac{\mathcal{K}_{\rm UV}}{\mathcal{K}_{\rm UV,0}}\right)^{-1}\left(\frac{\rm SFR_{\rm nuc}}{1.0~\msunyr}\right),
\end{equation}
where the conversion factor $\mathcal{K}_{\rm UV}$ depends on the stellar initial mass function (IMF), metallicity, and stellar age.
For a constant SFR and a Salpeter IMF ranging from $0.1$ to $100~\msun$, the value at $1500~{\rm \AA}$ asymptotically approaches
$\mathcal{K}_{\rm UV} \simeq 2\times 10^{-10}\msunyr \lsun^{-1}$ for ages $\gtrsim 300~{\rm Myr}$, and is only weakly dependent on metallicity for 
$Z=\zsun$ to $\zsun/50$ \citep{Madau_Dickinson_2014}.
Switching to a Kroupa or Chabrier IMF reduces $\mathcal{K}_{\rm UV}$ by $30-40\%$.
A top-heavy IMF (e.g., $50-500~\msun$) for metal-free stars ($Z=0$) increases the UV output per unit SFR, 
yielding $\mathcal{K}_{\rm UV} \simeq 0.54 \times 10^{-10}\msunyr\lsun^{-1}$ at $t \gtrsim 10~\mathrm{Myr}$ 
\citep{Zackrisson_2011,Inayoshi_2022c,Harikane_2024_galaxy}.
Adopting $\mathcal{K}_{\rm UV,0}= 10^{-10}~\msunyr\lsun^{-1}$ as our fiducial value at $t\gtrsim 10~{\rm Myr}$,
the UV luminosity becomes as high as $L_{\rm UV}\gtrsim 4\times 10^{43}~{\rm erg~s}^{-1}$, comparable to typical UV luminosities (uncorrected for dust)
of observed LRDs.
Note that the nuclear starburst discussed here shares similar properties with extremely metal-poor galaxies in the local universe, characterized by $Z\lesssim 0.1~\zsun$, stellar masses of $M_\star \simeq 10^5-10^7~\msun$, and ages of $4-50~{\rm Myr}$
\citep[e.g.,][]{Jerabkova_2017,Kojima_2020}.

In Figure~\ref{fig:z_growth}, we mark the halo conditions for ${\rm SFR}_{\rm nuc}\gtrsim 1.0~\msunyr$ (red thick curves).
This level of nuclear activity reproduces the UV luminosity observed in LRDs, $L_{\rm UV}\gtrsim 10^{10}~\lsun$. 
Additionally, we consider that the SFR decreases when the halo mass exceed the shock heating condition, $M_{\rm h}\gtrsim 10^{12}~\msun$ regardless of stellar and/or AGN feedback.
For quasar-host galaxies (left), the condition is satisfied at $6\lesssim z \lesssim 11$,
whereas for less massive halos (right), the corresponding redshift window shifts to $3\lesssim z \lesssim 8$.

\subsection{Composite V-shaped SED of LRDs}\label{sec:VSED}

Based on the energetics of the red and blue components in observed LRD spectra, we consider a model for the ``V-shaped" SEDs that distinguish LRDs from typical AGNs.
The optical component is attributed to an optically thick, gaseous envelope surrounding the BH, characterized with
a surface temperature $T_{\rm eff} \simeq 4000-5000~\K$.
In this model, the Wien tail of the blackbody spectrum naturally reproduces the red continuum without requiring additional dust reddening.
The blue UV component is modeled as emission from a young stellar population formed in the dense nuclear disk, in which a massive seed BH had already formed.
We adopt SED templates for low-metallicity galaxies ($Z = Z_\odot/50$) with a constant SFR and stellar ages 
of $10-100$~Myr \citep{Inoue_2011}, assuming a Salpeter IMF over $1-100~\msun$.
Nebular emission is included under the assumption that all ionizing photons beyond the Lyman limit are absorbed and reprocessed into lines (i.e., escape fraction $f_{\rm esc} = 0$), thereby maximizing line luminosities.
With a constant SFR, the stellar mass is the product of the SFR and stellar age. 
Since the UV continuum is dominated by short-lived massive stars ($t_{\rm age}\simeq 3-10~{\rm Myr}$), 
the UV luminosity per unit SFR saturates at $t\gtrsim {\rm a~few}\times 10~{\rm Myr}$.
In contrast, the optical luminosity increases with stellar age owing to the accumulation of lower-mass stars in the nuclear region.

In the left panel of Figure~\ref{fig:sed_fid}, we present the model SED for a LRD (black), consisting of two components: a BH envelope with $T_{\rm eff}=4300~\K$ radiating at the Eddington luminosity of a $M_{\rm BH}=5\times 10^5~\msun$ BH (red), and
a 10~Myr-old young starburst with ${\rm SFR}=1.0~\msunyr$ (blue).
For comparison, we overlay the average photometric SED of $z\sim 6$ LRDs from \citet{Akins_2025}, 
covering HTS, JWST/NIRCam and JWST/MIRI data.
We note that the HST~F814W and JWST~F090W filters probe wavelengths blueward and redward of the redshifted Ly$\alpha$ emission line, 
respectively, so that the Lyman break feature is not sharply resolved in the stacked photometry.

The observed red optical continuum is well reproduced by the a single-temperature blackbody spectrum with a Balmer break naturally imprinted by atmospheric radiative transfer in the BH envelope 
\citep[see also][]{Inayoshi_Maiolino_2025,Kido_2025,Liu_2025b,Nandal_Loeb_2025}.
The blue UV continuum is consistent with a young starburst with a stellar mass of $M_{\rm \star,nuc}=10^7~\msun$.
This stellar population produces a UV luminosity of $\nu L_\nu \simeq 4\times 10^{43}~{\rm erg~s}^{-1}$, 
corresponding to a high ${\rm SFR}_{\rm nuc} \simeq 1.0~\msunyr$, as discussed in Section~\ref{sec:OBUV}.
If the envelope luminosity indeed matches the Eddington luminosity of the central BH, the BH-to-stellar mass ratio at this early stage is expected to be $M_{\rm BH} / M_{\rm \star, nuc} \simeq 0.05$,
which is significantly higher than in typical low-redshift galaxies \citep[e.g.,][]{Kormendy_Ho_2013,Reines_Volonteri_2015}.

In the right panel of Figure~\ref{fig:sed_fid}, we show model SEDs for a LRD with a mass ratio of 
$M_{\rm BH}/M_{\rm \star,nuc}\simeq 0.27$, $0.15$, and $0.08$, where the envelope luminosity is assumed 
to be equivalent to the BH Eddington luminosity and the stellar age is 10 Myr.
Here, we apply moderate dust attenuation with visual extinction $A_V=0.1~{\rm mag}$, assuming 
the Small Magellanic Cloud extinction law.
For comparison, we overlay the NIRSpec MSA/PRISM spectra of several LRDs that exhibit a strong Balmer break feature:
A2744-QSO1 at $z = 7.04$ \citep{Furtak_2024,Ji_2025}, MoM-BH$^\ast$-1 at $z = 7.76$ \citep{Naidu_2025},
RUBIES-UDS-154183 at $z = 3.55$ \citep{deGraaff_2025b}, UNCOVER-45924 at $z = 4.47$ \citep{Labbe_2024b},
and CAPERS-LRD-z9 at $z=9.288$ \citep{Taylor_2025b}.
The luminosity is normalized at rest-frame $0.51~\mum$ for all the objects.
This composite SED model can reproduce the overall spectral shape of individual LRDs as well.
In this model framework, these LRDs with a strong Balmer break tend to require a higher mass ratio of 
$M_{\rm BH}/M_{\rm \star,nuc}$, compared to that constrained from the stacking LRD SED.

V-shaped SED models have been investigated in several recent studies with different proposed origins for the blue UV and red optical components.
In one-component scenarios, the uniformity of the SED is attributed to relatively well-motivated physical mechanisms using a small number of model parameters.
For instance, \cite{Inayoshi_2025b} considered accretion onto a binary BH system with a circum-binary disk and two mini-disks around each of the two BHs.
When the binary separation is $\sim 10^{3}$ Schwarzschild radii (for the total mass of the BHs), the UV and optical continua arise from the hotter mini-disks and colder circum-binary disk, respectively, while the turnover wavelength of the V-shaped SED reflects the deficit of thermal emission caused by gap opening due to binary torques.
Alternatively, \citet{C.Zhang_2025} proposed a self-gravitating disk configuration, where a similar SED shape emerges due to structure variations of the disk thermal profile.
Two-component models provide another route to reproducing the observed V-shaped SEDs of LRDs either by tuning physical parameters or through conditions seen in cosmological simulations \citep{Zwick_2025,Cenci_Habouzit_2025,Jeon_2025}.
However, the two-component scenarios face a key unresolved issue: explaining the remarkable uniformity of LRD SEDs and understanding its physical origin and redshift evolution.
Addressing this issue is the main focus in the following sections.

\vspace{2mm}

\section{Uniformity of the observed LRD V-shaped spectra}\label{sec:LRDend}

\subsection{Supernova feedback}\label{sec:SN}

As a natural consequence after the onset of starbursts, massive stars undergo core-collapse SN explosions (SNe) in the nuclear region, 
which inject mass, energy, and momentum into the surrounding medium.
We adopt the stellar IMF adopted in Section~\ref{sec:OBUV} (a Salpeter IMF with $1-100~\msun$) and assume that stars with masses of $\geq 10~\msun$ undergo SN explosions at the end of their stellar lifetime.
Under a constant SFR, the SN rate approaches a constant rate of ${\rm SFR}/\bar{m}_{\rm SN}$, where every $\bar{m}_{\rm SN}(=100~\msun)$ of stars produce one SN.
However, the rate is substantially lower in early stages of $<10-20~{\rm Myr}$, when a majority of massive stars are still alive.
We calculate the delay effect of the SN onset and give a fitting function of 
\begin{equation}
m_{\rm SN}(t) \equiv \bar{m}_{\rm SN}f(t) =\bar{m}_{\rm SN} \cdot\frac{1+(t_1/1.5)^{2.5}}{(t_1/1.5)^{2.5}},
\end{equation}
where $t$ is the elapsed time since the onset of star formation and $t_1=t/10~{\rm Myr}$.
Therefore, the cumulative number of SNe is approximated as \footnote{We calculate $m_{\rm SN}(t)$ using the simulated data of the stellar lifetime measured at the end of hydrogen burning for non-rotating stars with $Z=0.7~\zsun$ \citep{Ekstrom_2012}. Although a stellar population with $\zsun/50$ is considered as our fiducial case, the metallicity dependence of stellar age is moderate for massive stars with $\geq 10~\msun$ ($\lesssim 1\%$ difference between cases of $0.7~\zsun$ and $0.3~\zsun$; see \citealt{Eggenberger_2021}).}
\begin{equation}
    N_{\rm SN}=\frac{10^5}{f(t)} \cdot t_1 \bar{m}_{\rm SN,2}^{-1}\left(\frac{{\rm SFR}_{\rm nuc}}{\msunyr}\right).
\end{equation}
From Equation~(\ref{eq:KO15}), the total radial momentum injection into the surrounding gas in nuclear region is 
\begin{align}
    p_{\rm SNR}^{\rm tot} & \simeq 1.62~p_{\rm sf}N_{\rm SN},\\[2pt]
    & \simeq 1.8\times 10^{10}~\msun \cdot \kms N_{\rm SN,5},\nonumber
    \label{eq:KO15}
\end{align}
where $N_{\rm SN,5}=N_{\rm SN}/10^5$.
The simple estimate based on the product of the momentum injection and the cumulative number of SNe might overestimate the impact of SN feedback. In reality, some SN remnants could be removed from the nuclear region due to accretion onto the BH before the cumulative number reaches $\sim 10^5$, and collisions among multiple SN bubbles may dissipate part of the outward linear momentum.

Next, we examine the conditions where the total momentum injected by successive SNe in the nuclear region becomes strong enough to blow gas from the nuclear and disk scale ($R_{\rm nuc}=0.1~R_{\rm d}$).
The gaseous disk is bound by the gravitational potential of gas itself, dark matter, and the central BH.
The velocity dispersion of gas in the disk is calculated as 
\begin{align}
    \delta v_{\rm d}& \simeq \sqrt{\frac{G M_{\rm d}}{R_{\rm d}}}= \sqrt{\frac{f_{\rm b}f_{\rm d}}{\lambda_{\rm dm}}}V_{\rm circ},\nonumber\\
    & \simeq 163~\kms f_{\rm d,0.3}^{1/2}\bar{\lambda}^{-1/2}T_{\rm vir,6}^{1/2}.
\end{align}
We note that the velocity dispersion is approximated as a constant value over $r \lesssim R_{\rm d}$, 
but the value increases inside the BH gravitational sphere of influence defined by $R_{\rm inf}=GM_{\rm BH}/(\delta v_{\rm d})^2 \simeq 0.16~M_6~\pc \ll R_{\rm nuc}$. 
Therefore, the critical disk mass when momentum feedback begins to operate is given by $M_{\rm d,crit}\equiv p_{\rm SNR}^{\rm tot}/\delta v_{\rm d}$ or
\begin{align}
    M_{\rm d,crit} 
    \simeq 1.1\times 10^{8}~\msun~N_{\rm SN,5} f_{\rm d,0.3}^{-1/2}\bar{\lambda}^{1/2}T_{\rm vir,6}^{-1/2}.
\label{eq:Md,critSN}
\end{align}
This value is comparable to the disk mass in the nuclear region, and therefore the nuclear region would be evacuated when more than $N_{\rm SN}\gtrsim 10^5$ SNe occur.
The condition for $M_{\rm d}(<R_{\rm nuc})<M_{\rm d,crit}$ is further rewritten as
\begin{equation}
    \frac{t_1}{f(t_1)} > 0.90 ~\psi T_{\rm vir,6}^{1/2},
\end{equation}
where $\psi \equiv \epsilon_{-1}^{-1} \tau_{1}f_{\rm d,0.3}^{1/2} \bar{\lambda}^{-1/2}\bar{m}_{\rm SN,2}\simeq \mathcal{O}(1)$.
The condition depends only on the halo virial temperature, but not explicitly on redshift.
We also note that this condition is independent of $f_{\rm n}$ (i.e., the choice of $R_{\rm nuc}$) because both the disk mass 
and the number of SNe within the nuclear region are proportional to $f_{\rm n}$.

%%%%%%%%%%
%%% Fig. 4 %%%
%%%%%%%%%%
\begin{figure}
\centering
\includegraphics[width=84mm]{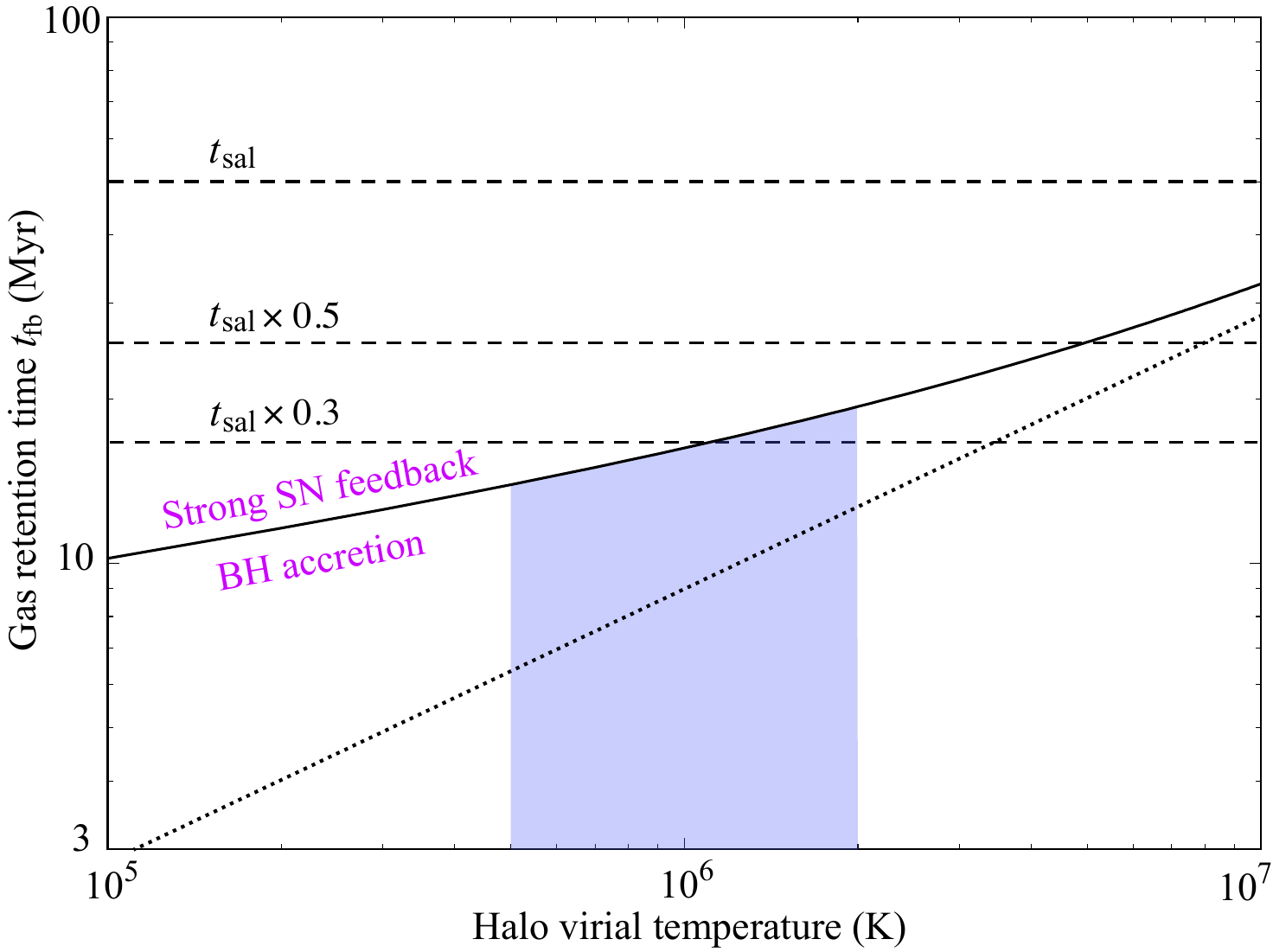}
\caption{Maximum time duration (solid curve) over which gas can be retained 
in the nuclear region against momentum feedback from multiple SNe, as a function of halo virial temperature.
The dotted curve indicates the gas retention time obtained without accounting for the time delay of SN explosions (see text).
For massive halos with $T_{\rm vir,6}=0.5-2$ (shaded region), the gas retention time reaches $t_{\rm fb}\simeq 14-19~{\rm Myr}$, 
corresponding to $\sim 30-40\%$ of a Salpeter timescale.
}
\label{fig:SNtime}
\vspace{2mm}
\end{figure}

%%%%%%%%%%
%%% Fig. 5 %%%
%%%%%%%%%%
\begin{figure*}
\centering
\includegraphics[width=87mm]{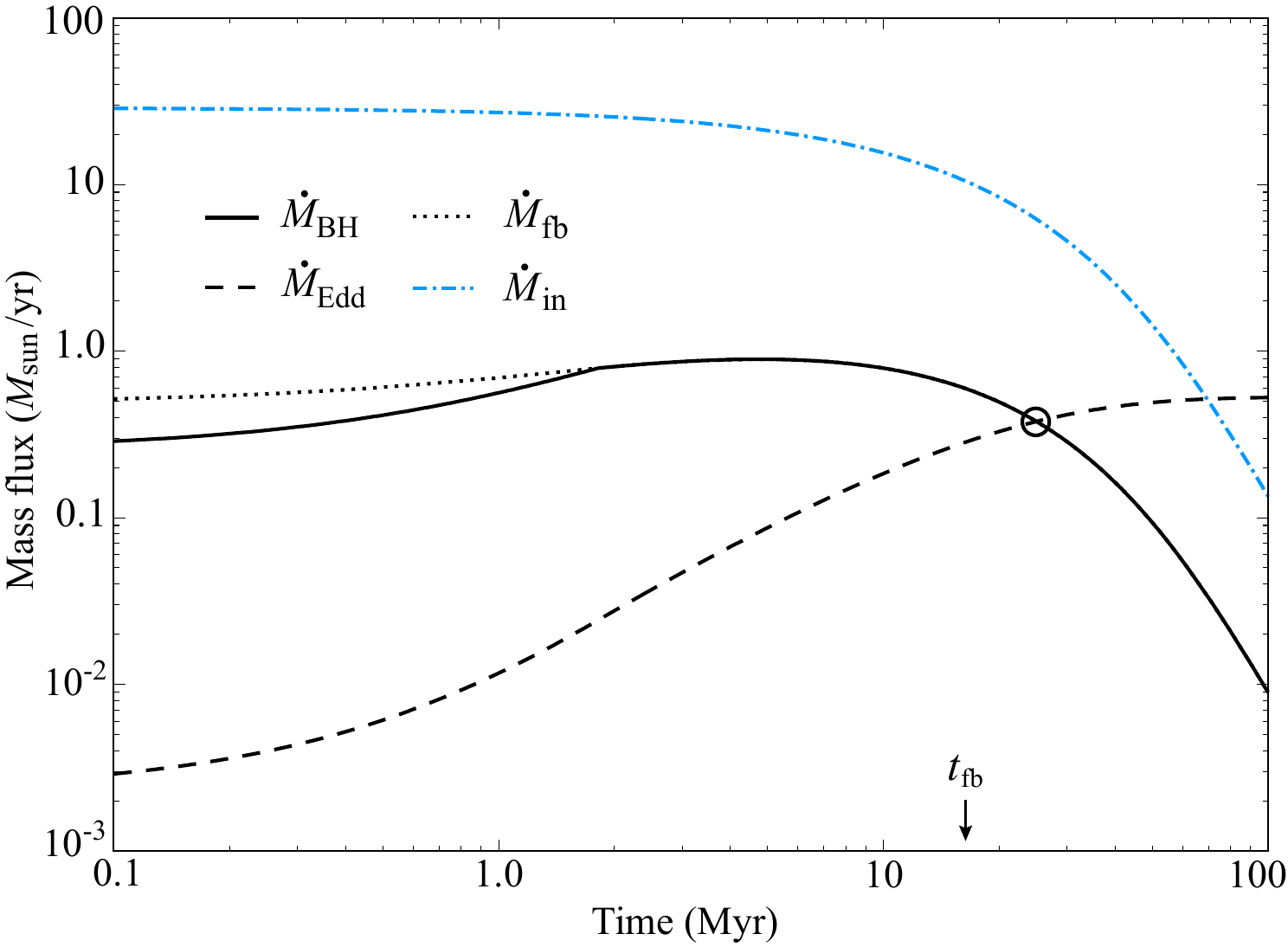}\hspace{5mm}
\includegraphics[width=87mm]{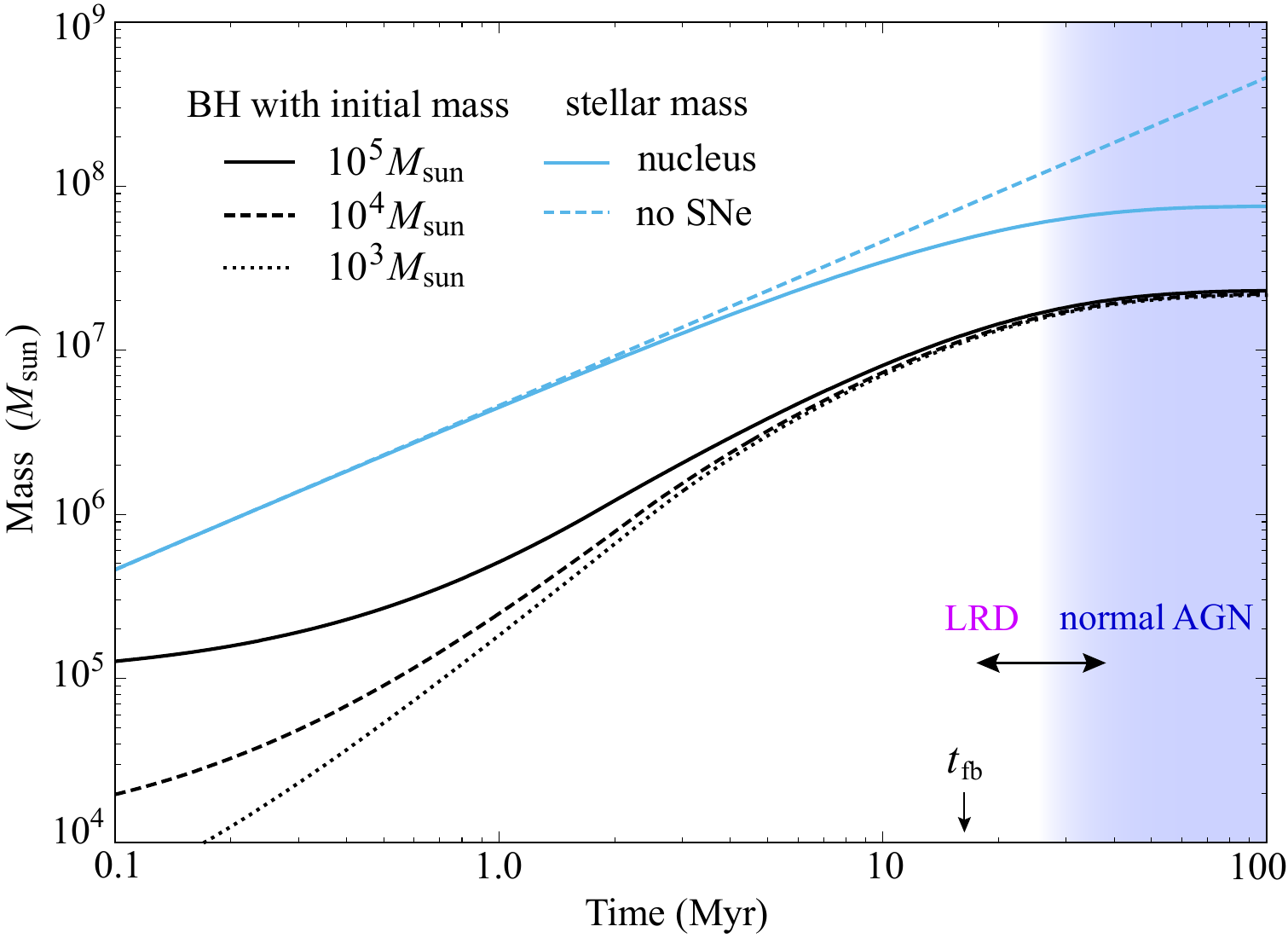}
\caption{Time evolution of BHs and stars in a massive halo taken as our fiducial case ($T_{\rm vir}=10^6~\K$ and $1+z=10$).
{\it Left}: the inflow rate on the envelope ($\dot{M}_{\rm in}$),
the BH feeding rate ($\dot{M}_{\rm BH}$), the rate regulated by BH momentum feedback ($\dot{M}_{\rm fb}$), and the Eddington rate ($\dot{M}_{\rm Edd}$).
The initial BH mass is set to $10^5~\msun$ and $\epsilon_{\rm BH}=0.2$ is adopted.
At later times ($t\gtrsim 20~{\rm Myr}$), the BH feeding rate falls below the Eddington limit, the envelope structure is no longer maintained, and thus the LRD phase terminates.
{\it Right}:
Black curves present BH masses with three different initial masses of $M_{\rm BH,0}=10^3$ (dotted), $10^4$ (dashed), and $10^5~\msun$ (solid) for $\epsilon_{\rm BH}=0.2$.
Cyan curves show the nuclear mass with SN feedback (solid) and without SN feedback (dashed).
By the time when the LRD phase ends, when the BH feeding rate falls below the Eddington limit, 
the BH mass reaches $M_{\rm crit,LRD}\simeq 10^7~\msun$.
}
\label{fig:BHgrowth}
\vspace{5mm}
\end{figure*}

Figure~\ref{fig:SNtime} shows the maximum time duration (solid curve) over which gas can be retained in the nuclear region against momentum injection from multiple SNe, as a function of halo virial temperature.
The gas retention time $t_{\rm fb}$ increases monotonically with halo virial temperature,
For comparison, the dotted curve indicates the maximum duration $t_{\rm fb}$ obtained without accounting for the time delay of SN explosions (i.e., assuming $m_{\rm SN}(t)=\bar{m}_{\rm SN}$).
The correction for the delay effect becomes important in lower-mass halos, where it extends $t_{\rm fb}$ significantly.
For massive halos with $T_{\rm vir,6}=0.5-2$, the maximum retention time reaches $t_{\rm fb}\simeq 14-19~{\rm Myr}$, corresponding to $\sim 30-40\%$ of a Salpeter timescale.

\subsection{Termination of LRD phases}\label{sec:AGNfeedback}

As discussed in Section~\ref{sec:BHenv}, the envelope structure is maintained by strong ram pressure of inflowing gas 
onto the surface against momentum feedback via super-Eddington disk winds (also see \citealt{Kido_2025}).
The momentum injection rate from the accreting BH to the envelope is proportional to the BH feeding rate $\dot{M}_{\rm BH}$ as 
\begin{equation}
    \dot{P}_{\rm out} = \eta_{\rm w} c \dot{M}_{\rm BH},
\end{equation}
where $\eta_{\rm w}$ is the outflow efficiency.
Recent radiation-hydrodynamic simulations for BH accretion find that the outflow efficiency settles down to
$\eta_{\rm w}\sim 0.03$ in a quasi-steady state of super-Eddington flows \citep{Hu_2022a}.
We note that the outflow efficiency would increase to $\eta_{\rm w} \simeq \mathcal{O}(0.1)$
when the outflow traps radiation and becomes adiabatic, where most the energy is transferred to the kinetic energy of the outflow.
Thus, we adopt $\eta_{\rm w}=0.1$ in the following analysis, but keep the dependence on the efficiency.
In contrast, the inflow momentum rate onto the envelope surface is 
\begin{align}
    \dot{P}_{\rm in} = v_{\rm ff} \dot{M}_{\rm in},
\end{align}
where $v_{\rm ff}$ is the inflow velocity and $\dot{M_{\rm in}}$ is the mass inflow rate onto the envelope,
which is assumed to be a fraction $\epsilon_{\rm BH}$ of the baryonic inflow rate in Equation~(\ref{eq:Mdotb2}), $\dot{M}_{\rm in}=\epsilon_{\rm BH}\dot{M}_{\rm b} e^{-t/t_{\rm fb}}$.
Here, the $e^{-t/t_{\rm fb}}$ term reflects suppression of gas supply owing to SN feedback taking place in the nuclear region.

To satisfy $\dot{P}_{\rm out}/ \dot{P}_{\rm in}\lesssim 1$, for a given inflow rate onto the envelope surface, the BH feeding rate is limited to 
\begin{align}
\dot{M}_{\rm BH}&\lesssim \dot{M}_{\rm fb}\equiv \frac{v_{\rm ff}}{\eta_{\rm w} c}\dot{M}_{\rm in}, \nonumber\\
    &\simeq 0.03~\dot{M}_{\rm in}\eta_{-1}^{-1}\lambda_{\rm Edd}^{-1/4}M_6^{1/4}T_0,
\end{align}
where $v_{\rm ff}=V_{\rm ph}$ is set so that the response of the envelope radius caused by BH mass growth is considered. 
When the BH feeding rate is lower than the critical value, we adopt a scaling relationship 
of $\dot{m}_{\rm BH}\simeq \dot{m}_{\rm in}^{1/2}$ obtained for super-Eddington accreting BHs,
where $\dot{m}$ denotes the Eddington ratio of the mass flux \citep{Hu_2022a}.
Therefore, the BH feeding rate is given as 
\begin{equation}
 \dot{M}_{\rm BH} =  {\rm min}\left( \dot{M}_{\rm fb},  
 ~\dot{m}_{\rm in}^{1/2} \dot{M}_{\rm Edd}\right),
\end{equation}
which is used when $\dot{M}_{\rm BH}\gtrsim \dot{M}_{\rm Edd}$.

The left panel of Figure~\ref{fig:BHgrowth} presents the evolution of BH growth rate 
in a massive halo with $T_{\rm vir}=10^6~\K$ at $1+z=10$:
the mass inflow rate on the envelope ($\dot{M}_{\rm in}$),
the BH feeding rate ($\dot{M}_{\rm BH}$), the critical rate regulated by BH momentum feedback
($\dot{M}_{\rm fb}$), and the Eddington rate ($\dot{M}_{\rm Edd}$).
The initial BH mass is set to $10^5~\msun$ and $\epsilon_{\rm BH}=0.2$ is adopted.
At early times, the accretion rate onto the envelope is extremely super-Eddington
($\dot{m}_{\rm in}\simeq 10^4$), but most of this inflowing gas is recycled within the envelope through convective eddies\footnote{During the earliest stage, the envelope mass greatly exceeds the mass of the BH. As the BH grows via super-Eddington accretion and the system enters the LRD phase (see below), the two masses become comparable, as discussed in \citet{Kido_2025}.}.
As a result, the effective BH feeding rate is reduced to $\dot{M}_{\rm BH}\sim 0.3~\msunyr$,
though it is still super-Eddington ($\dot{m}_{\rm BH}\gtrsim 100$).
As the BH grows, the feeding rate becomes limited by momentum feedback,
but remains nearly constant until the inflow onto the envelope begins to decline due to 
gas removal by SN feedback in the nuclear region at $\sim 70~\pc$ 
($t\gtrsim t_{\rm fb}\simeq 16~{\rm Myr}$).
At later times ($t\gtrsim 25~{\rm Myr}$), the BH feeding rate falls below the Eddington limit, the envelope structure is no longer maintained, and thus the LRD phase terminates.

The right panel of Figure~\ref{fig:BHgrowth} shows the evolution of BH and stellar mass in the same halo.
For the stellar component (cyan curves), we present the nuclear stellar mass with SN feedback (solid) and without SN feedback (dashed).
We also compare three different initial BH masses of $M_{\rm BH,0}=10^{3}-10^{5}~\msun$ (black curves).
At early times, the BH masses reflect their initial values, but rapid super-Eddington accretion soon erases this memory. 
By $t\simeq 10~{\rm Myr}$, all cases converge to $M_{\rm BH}\simeq 8\times 10^6~\msun$, regardless of the initial mass.
By the time when the LRD phase ends, when the BH feeding rate falls below the Eddington limit, 
the BH mass reaches $\simeq 2\times 10^7~\msun$.
The BH mass measured at $t=t_{\rm fb}$ is derived by analytically solving $\dot{M}_{\rm BH}=\dot{M}_{\rm fb}$ as
\begin{align}
 M_{\rm crit,LRD}  \simeq 6.4\times 10^6~\msun ~ \xi t_{\rm fb,1}^{4/3} T_{\rm vir,6}^2\left(\frac{1+z}{10}\right)^{4/3},
 \label{eq:critLRD}
\end{align}
where $\xi = (\epsilon_{\rm BH,0.2}\eta_{-1}^{-1}\lambda_{\rm Edd}^{-1/4}T_0)^{4/3}$ and the initial BH mass is assumed to be negligible in this expression.
As a result of super-Eddington growth and its suppression due to SN feedback, 
the ratio of BH mass to nuclear stellar mass increases rapidly and converges to 
$\gtrsim 0.1$ by the end of LRD phase.
This convergence behavior plays a key role in shaping the characteristic V-shaped SED of LRDs (see Section~\ref{sec:Lratio}).
Figure~\ref{fig:z_growth} illustrates that this critical BH mass evolves with the assembly history of dark matter halos (green solid).
The final BH mass attainable through an LRD phase is determined by the onset redshift of super-Eddington accreting growth stages.
Once the BH mass reaches this threshold, further LRD activity is suppressed until halo conditions greatly change on cosmological timescales
and $M_{\rm crit,LRD}\gg M_{\rm BH}$ is re-established.

Intriguingly, this critical BH mass for maintaining the LRD phase against momentum feedback ($M_{\rm crit,LRD}$) is comparable to
the mass threshold for sustaining super-Eddington accretion 
against AGN photoionization and heating in massive halos \citep{Inayoshi_Haiman_Ostriker_2016},
\begin{equation}
 M_{\rm crit,ion}\simeq 1.4\times 10^7~\msun~T_{\rm vir,6}.
 \label{eq:crithyper}
\end{equation}
This condition is applicable when the emergent radiation spectrum is as UV bright as in typical AGNs without reprocessing 
to optical photons through a thick gaseous envelope.
In such cases, once the BH mass reaches the threshold ($M_{\rm crit,ion}$), the ionization front expands rapidly and 
the outward gas pressure within the ionized region pushes the surrounding gas, thereby suppressing mass accretion.
If the ionization front stalls through balance with radiative recombination, the depletion of hot gas creates
an inward pressure gradient that in turn triggers renewed mass inflows \citep[e.g.,][]{Milosavljevic_2009,Park_Ricotti_2011,Park_Ricotti_2012,Sugimura_2017,Ogata_2024}.  
This mechanism leads to episodic accretion cycles in which the time-averaged accretion rate rarely exceeds the Eddington limit,
while pre-existing super-Eddington accretion bursts would last for $\sim 10~{\rm Myr}$ even after the onset of the feedback process
\citep[e.g.,][]{Prole_2025}.
This second critical BH mass is presented in Figure~\ref{fig:z_growth} (green dashed).
For the halo conditions of interest, one finds $M_{\rm crit,ion}\gtrsim M_{\rm crit,LRD}$,
except in the quasar-host case at $z\lesssim 9$.
The coincidence of these two critical masses suggests that, even if a BH experiences an LRD phase at its seeding stage, once the BH mass exceeds the two thresholds, i.e., $M_{\rm BH}\gtrsim {\rm max}(M_{\rm crit,ion}, M_{\rm crit,LRD})$, the BH hardly re-enters another LRD phase.
Such a dramatic change in the accretion mode marks a point of no return, terminating the LRD phase and transitioning to a normal AGN phase.

Importantly, in both halo assembly cases, the BH mass reachable through LRD phases is limited to $\sim 4\times 10^7~\msun$ (left) and 
$\sim 10^7~\msun$ (right).
The red shaded region indicates where LRDs can emerge with characteristic V-shaped SEDs produced by optically-red envelopes powered by BH accretion and UV-bright young starbursts in the nuclear region.
At lower redshifts, some characteristic properties of LRDs are suppressed: either because the halo mass reaches the threshold for virial shock heating, or because the accumulated stellar mass has grown to $M_\star \gtrsim 10^{10}~\msun$, with a stellar size large enough to be resolved by JWST/NIRCam imaging \citep{Killi_2024,Rinaldi_2025,Chen_2025a,Chen_2025b,Billand_2025}.

%%%%%%%%%%
%%% Fig. 6 %%%
%%%%%%%%%%
\begin{figure*}
\centering
\includegraphics[width=118mm]{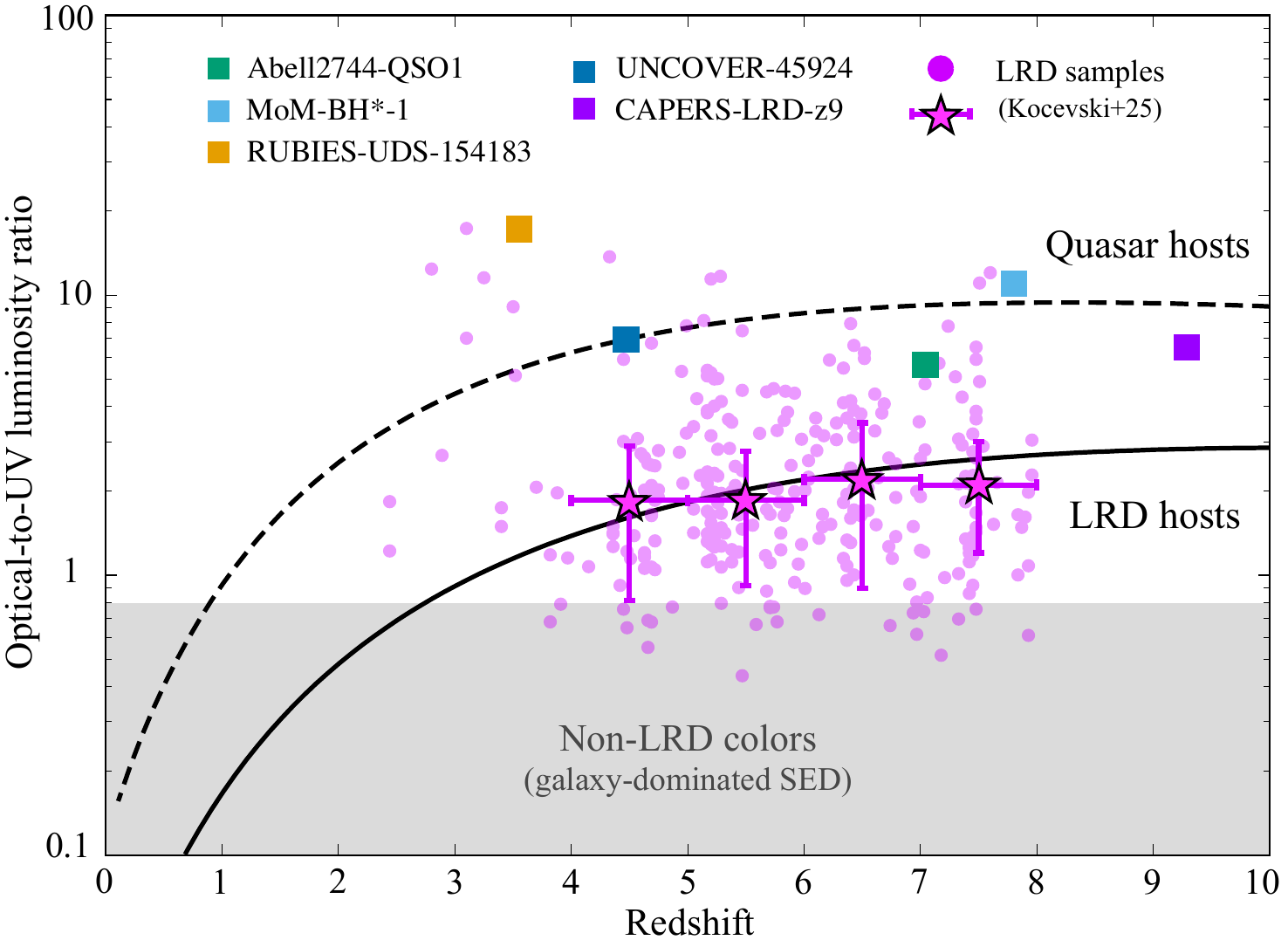}
\caption{Optical-to-UV luminosity ratio of LRDs calculated from the two component model with a BH-envelope (optical) and compact starbursts (UV).
The black curves present the model-predicted ratios for LRDs that reside in quasar-host galaxies (dashed) and less-massive galaxies (solid; denoted as LRD hosts).
Purple circles and stars indicate the values estimated from the LRD samples compiled by \citet{Kocevski_2025} and the median value at each redshift bin. 
Square symbols present those of a subset of the LRD sample with a deep Balmer break \citep[][see text]{Furtak_2024,Labbe_2024b,deGraaff_2025b,Naidu_2025,Taylor_2025b}.
The gray shaded region defined with $L_{\rm opt}/L_{\rm UV}\leq 0.8$ corresponds to colors that have $\beta_{\rm opt}\leq 0$ with $\beta_{\rm UV}=-2$, where the SED is dominated by a blue component and is not classified as an LRD.
}
\label{fig:L_ratio}
\vspace{2mm}
\end{figure*}

The upper limit on BH mass in LRD phases predicted by our model naturally sets the maximum luminosity at
\begin{equation}
    \lambda L_{\rm 5100,max} \simeq 2\times 10^{45}~{\rm erg~s}^{-1}~\lambda_{\rm Edd}
    \left(\frac{M_{\rm BH,max}}{4\times 10^7~\msun}\right),
\end{equation}
where the rest-frame 5100~\AA\ monochromatic luminosity is converted by assuming  
an envelope temperature of $T_{\rm eff}=5000~\K$ (the bolometric correction factor is 1.78).
This theoretical limit aligns well with the sharp cutoff in the optical luminosity function at $\lambda L_{5100}\simeq 2.5\times 10^{45}~{\rm erg~s}^{-1}$ \citep{Ma_2025b}.

Finally, we note a caveat of our argument.
We have focused on how BHs grow during an LRD phase, and how halo conditions {\it can} trigger this phase and 
determine the terminal BH mass through stellar feedback.
However, we do not specify how or where the LRD stage {\it actually} begins, since its onset depends on the seeding mechanism and the initial BH mass.
In our framework, as long as a seed BH resides in the dense nuclear region of a protogalaxy, it can undergo super-Eddington accretion.
Even in low-mass halos with $T_{\rm vir}\ll 10^6~\K$, the gas retention time $t_{\rm fb}$ cannot be shorter than the lifetime 
of massive stars ($\gtrsim 3~{\rm Myr}$), allowing for a short-lived LRD episode.
By contrast, when the seed mass is as large as $M_{\rm BH,0}\simeq 10^{4}-10^{5}~\msun$ (either due to heavy seed formation or prior growth), 
the available LRD phase in low-mass halos with $T_{\rm vir}\ll 10^5~\K$ becomes negligible, since the critical mass for momentum feedback 
is already comparable to the BH mass.
Therefore, LRD activity of interest is expected to emerge primarily in halos with $T_{\rm vir}\simeq 5\times 10^5-10^6~\K$, hosting BHs with 
$M_{\rm BH}\ll 10^6~\msun$ at the onset of the LRD.
Here, our aim is not to model the actual formation pathway of LRDs, but rather to investigate the physical properties of BHs 
once they enter this phase and to explain how they sustain the uniform V-shaped SEDs observed in LRDs.

\subsection{Stability of the V-shaped spectra}\label{sec:Lratio}

The characteristic V-shaped SED of LRDs appears consistent across different sources, 
and the optical-to-UV luminosity ratio ($\nu L_\nu$ in erg s$^{-1}$) falls within a relatively narrow range.
In Figure~\ref{fig:L_ratio}, we show the luminosity ratios for LRDs 
(purple circles for individual objects and stars with error bars for median values
in each redshift bin) compiled by \citet{Kocevski_2025}.
The ratios are derived from the apparent UV magnitude ($m_{\rm UV}$ at rest-frame 1450~\AA)
and the F444W magnitude ($m_{\rm F444W}$)\footnote{Flux data for each LRD sample are 
available at \citet{Kocevski_2025}.}.
Since the F444W flux density does not fully trace the optical emission of LRDs at $z \gtrsim 6-7$, 
the plotted data likely underestimate the optical luminosity and extrapolation of the SED data to the redder side of optical bands ($\simeq 1~\mum$ in rest frame) would shift the ratios upward. 
Despite moderate scatter, the observed ratios remain narrowly distributed with median values $L_{\rm opt}/L_{\rm UV}\simeq 1.85-2.20$ and maxima reaching $L_{\rm opt}/L_{\rm UV}\simeq 10$.
Additionally, we overlay a rarer but non-negligible subset of the LRD samples with prominent Balmer break features \citep[square symbols;][]{Furtak_2024,Labbe_2024b,deGraaff_2025b,Naidu_2025,Taylor_2025b}\footnote{The luminosity ratio is derived from the photometric flux ratio between F150W and F770W for MoM-BH$^\ast$-1 and F150W and F1000W for CAPERS-LRD-z9.
For Abell2744-QSO1, where MIRI data are currently unavailable, we estimate the optical luminosity by extrapolating the F444W flux density to rest-frame $7000~{\rm \AA}$, assuming an optical slope of $\beta_{\rm opt}=1.2$.
For RUBIES-UDS-154183, the UV luminosity is obtained by extrapolating the F090W flux density to rest-frame $1450~{\rm \AA}$, adopting a UV slope of $\beta_{\rm UV}=-2.0$. If the UV slope of this object is steeper, the optical-to-UV luminosity ratio decreases significantly.}.
The gray shaded region defined with $L_{\rm opt}/L_{\rm UV}\leq 0.8$ corresponds to colors that have $\beta_{\rm opt}\leq 0$ with $\beta_{\rm UV}=-2$.
Within the region, the SED is dominated by a blue component and is not classified as an LRD \citep[e.g.,][]{Kocevski_2025,Hainline_2025,Hviding_2025}.

%%%%%%%%%%
%%% Fig. 7 %%%
%%%%%%%%%%
\begin{figure}
\centering
\includegraphics[width=83mm]{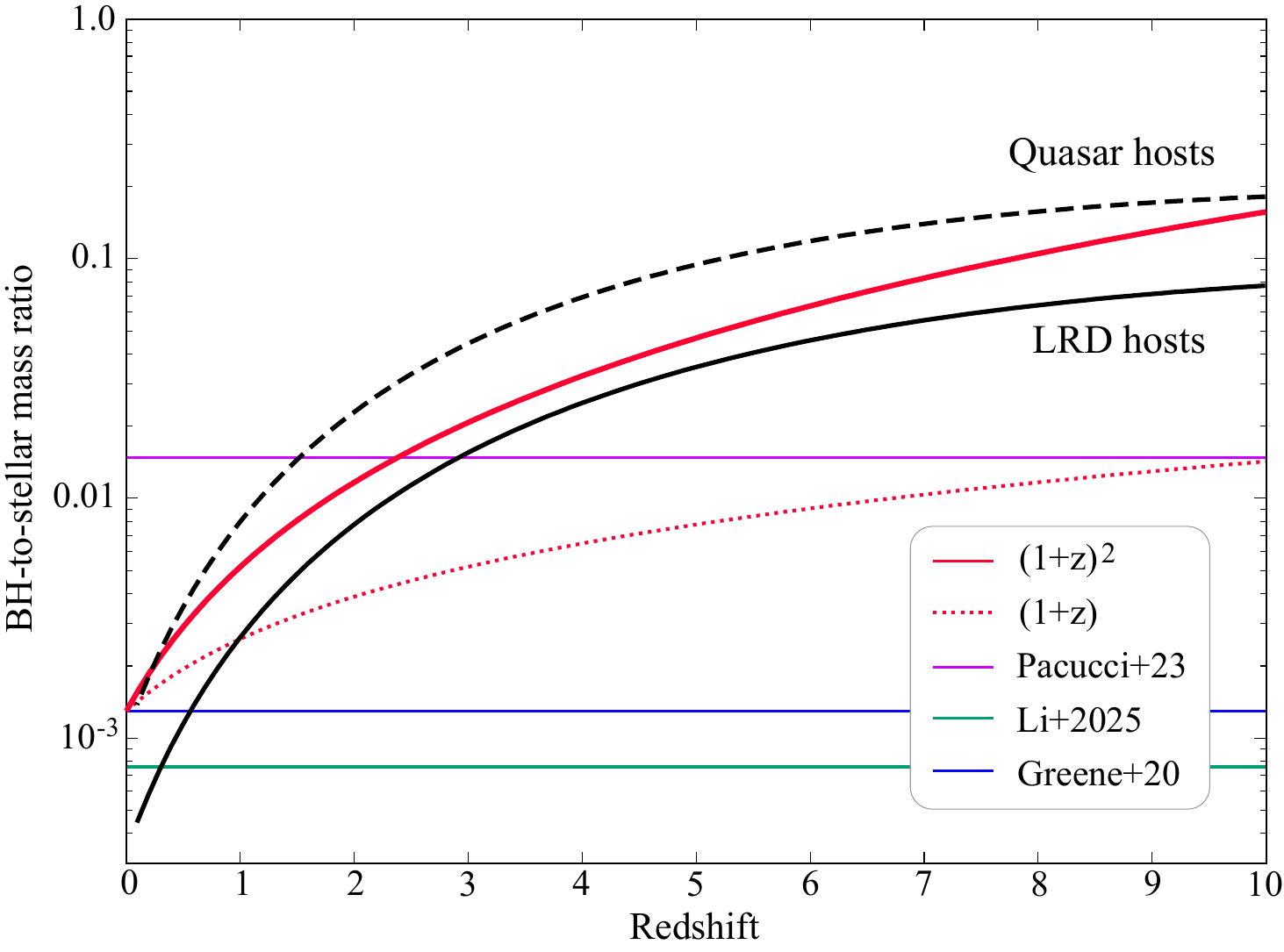}
\caption{BH-to-stellar mass ratio reachable during an LRD phase before SN feedback operates. 
The black curves show the model predictions for quasar-host galaxies (dashed) and less massive galaxies (solid).
The horizontal lines indicate the $M_{\rm BH}/M_\star$ ratio estimated for high-$z$ AGNs 
with JWST \citep{Pacucci_2023,J.Li_2025a} and for local SMBHs \citep{Greene_ARAA_2020}.
The red curves illustrate simple extrapolation from the local value as 
$\propto (1+z)^2$ and $(1+z)$.}
\label{fig:M_ratio}
\vspace{2mm}
\end{figure}

%%%%%%%%%%
%%% Fig. 8 %%%
%%%%%%%%%%
\begin{figure*}
\centering
\includegraphics[width=116mm]{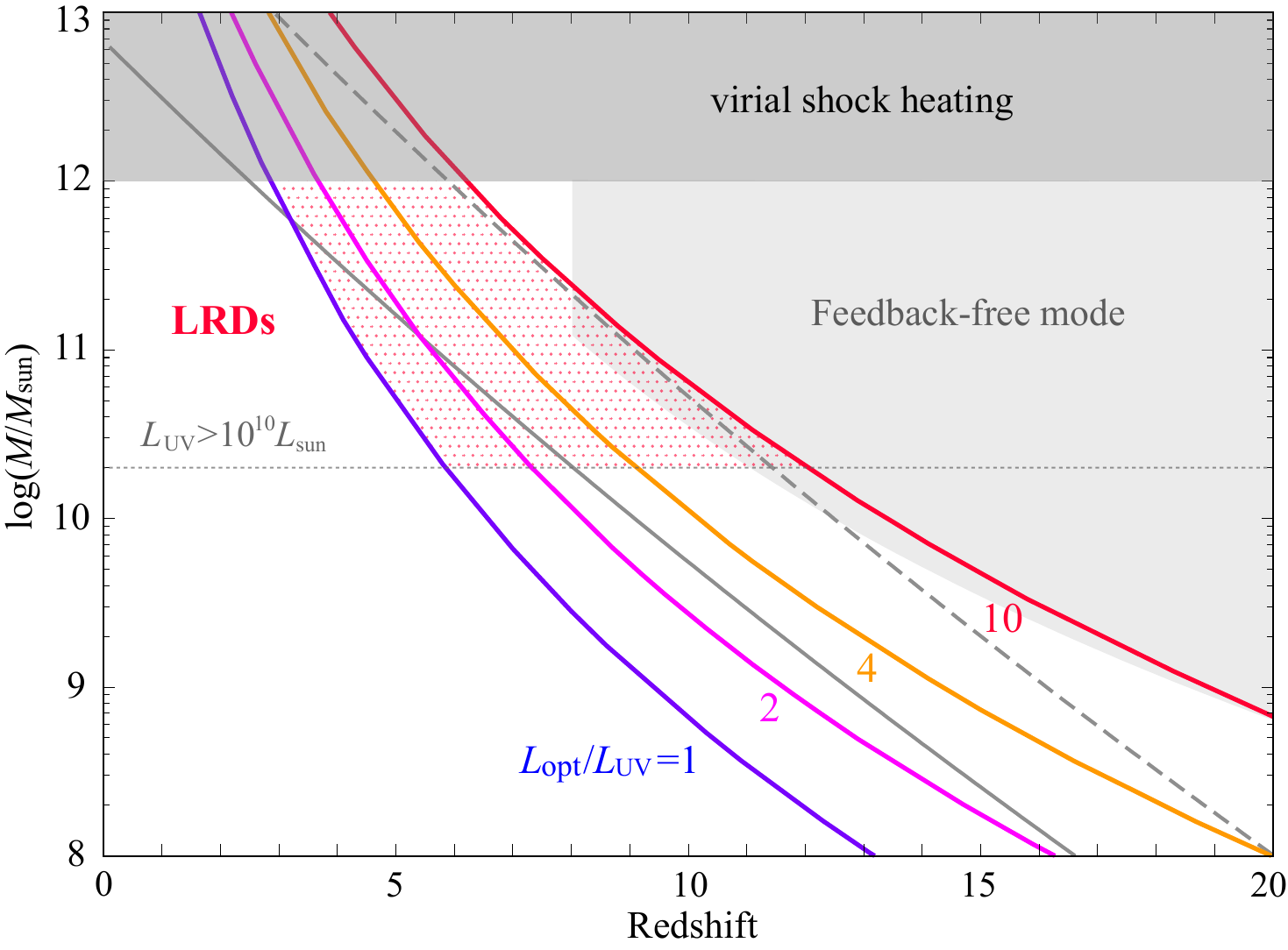}
\caption{Halo conditions leading to LRD phases. Each solid curve shows the optical-to-UV luminosity ratio at different halo conditions where a LRD emerges: 
$L_{\rm opt}/L_{\rm UV}=1$ (LRD color boundary), $2$ (typical LRDs), $4$, and $10$ (extremely red LRDs). 
The gray curves indicate the assembly histories of quasar-host galaxies (dashed) and less-massive galaxies (solid).
Overall, as the halo mass increases toward lower redshifts, 
the optical-to-UV luminosity ratio decreases from an nearly constant value.
The dotted line and dark-gray shaded region mark halo masses of $M_{\rm h}\simeq 2\times 10^{10}~\msun$ and $M_{\rm h}=10^{12}~\msun$, between which nuclear starbursts produce UV luminosities of $L_{\rm UV}\gtrsim 10^{10}~\lsun$.
The region filled with red dots corresponds to halo conditions where the BH envelope and starburst together produce the characteristic V-shaped SEDs of LRDs ($1\lesssim L_{\rm opt}/L_{\rm UV}\lesssim 10$).
The light-gray shaded region highlights the onset of feedback free starbursts \citep{Dekel_2023}, where significantly redder SEDs dominated by the BH-envelope emerge, though such cases are rarer.
}
\label{fig:LRD_condition}
\vspace{2mm}
\end{figure*}

Figure~\ref{fig:L_ratio} also shows the optical-to-UV luminosity ratio predicted based on the two-component model described in this paper.
We assume that the LRD phase persists for the stellar feedback timescale at each redshift, and evaluate luminosities at $t = t_{\rm fb}$ from the onset of star formation.
Although the luminosity ratio is time dependent, proto-LRDs in the earliest stages when the BH mass is still low would show lower optical-to-UV ratios. 
However, these young phases are short-lived, and observations likely capture the converged ratios maintained during the long-lived stage.
This treatment is justified as long as the activity duration is shorter than the cosmic interval 
corresponding to a significant redshift evolution ($\Delta z \simeq 1$).
The converged luminosity ratio is shown at each redshift in the cases where LRDs emerge in the quasar-host halo (dashed) and in less massive halo (solid).
In both cases, though the halo mass (and virial temperature) increases along the growth tracks as shown 
in Figure~\ref{fig:z_growth}, the optical-to-UV luminosity ratio remains nearly constant at high redshifts ($4<z<10$), owing to simultaneous SN feedback onto the two component.
The less-massive halo case reproduces the observed ratios of the majority of LRDs with 
$L_{\rm opt}/L_{\rm UV}\simeq 2$ \citep[purple circles and stars;][]{Kocevski_2025}.
Toward lower redshifts, the optical-to-UV luminosity ratio rapidly declines because of the saturation of BH masses due to AGN feedback.
The composite spectra are not classified as LRDs at $z\lesssim 2.5$.

On the other hand, the quasar-host case predicts higher ratios ($L_{\rm opt}/L_{\rm UV}\simeq 10$) consistent with a rarer but non-negligible subset of the LRD sample
\citep{Furtak_2024,Labbe_2024b,deGraaff_2025b,Naidu_2025,Taylor_2025b}.
Notably, several LRDs including RUBIES-UDS-154183 exhibit luminosity ratios of $\gtrsim 10$. These high ratios can be explained for $\xi\gtrsim 2-3$, corresponding to $\epsilon_{\rm BH}\gtrsim 0.4$ or $\eta_{\rm w}\lesssim 0.05$.

The luminosity ratio and its redshift dependence can be understood as follows.
In the cases of interest, where the BH mass growth is determined by momentum feedback ($M_{\rm crit,LRD}$),
the luminosity ratio between the optical continuum from the BH envelope and the UV continuum from the nuclear starburst is calculated as
\begin{align}
\mathscr{L}&\simeq 4.9~ \xi \zeta^{-1}
\lambda_{\rm Edd}t_{\rm fb,1}^{4/3}
 T_{\rm vir,6}^{1/2}\left(\frac{1+z}{10}\right)^{17/6},\nonumber\\[3pt]
 & \propto (1+z)^{17/6-3\omega/4}\longrightarrow \left(\frac{1+z}{10}\right)^{0.58},
 \label{eq:L_ratio}
\end{align}
where $t_{\rm fb}\propto T_{\rm vir}^{3/16}$ is approximated over $1 \lesssim T_{\rm vir,6} \lesssim 3$, 
and the power-law index $\omega$ is evaluated as
\begin{equation}
\omega \equiv -\frac{{\rm d}\ln T_{\rm vir}}{{\rm d}\ln z}=-1+\frac{2k_{\rm h}}{3}(1+z),
\end{equation}
assuming a halo assembly history (i.e., $M_{\rm h}\propto e^{-k_{\rm h}z}$).
For instance, with $k_{\rm h}=0.6$ at $1+z=10$, one obtains $\omega\simeq 3$ and thus $\mathscr{L}\propto (1+z)^{0.58}$.
This weak redshift dependence remains the ratio constant at high redshifts, as shown in Figure~\ref{fig:L_ratio}.
At lower redshifts ($z\lesssim 4$), the slope steepens to $\mathscr{L}\propto (1+z)^2$ and thus the ratio decreases rapidly below the LRD color selection threshold.

Figure~\ref{fig:M_ratio} presents the ratio between the BH mass ($M_{\rm crit,LRD}$) and nuclear 
stellar mass ($M_{\rm \star,nuc}$) at the end of an LRD phase.
The black curves are the model predictions for quasar-host galaxies (dashed) and less massive galaxies (solid).
The mass ratio inferred for high-$z$ AGNs with JWST \citep{Pacucci_2023,J.Li_2025a} and for local SMBHs \citep{Greene_ARAA_2020} are overlaid for comparison.
The red curves illustrate simple extrapolation from the local value as $\propto (1+z)^2$
and $\propto (1+z)$.
The predicted mass ratio falls within a relatively narrow rage at each redshift 
($M_{\rm crit,LRD}/M_{\rm \star,nuc}\simeq 0.1$ at $5<z<10$).
At lower redshifts ($z\lesssim 1$), the mass ratio becomes increasingly dominated by the stellar component and eventually approaches the local value of $\sim 0.1\%$ 
\citep{Kormendy_Ho_2013,Reines_Volonteri_2015,Greene_ARAA_2020}; see also a potential evolution of the scaling relation from low-luminosity to high-luminosity AGNs 
including LRDs and unobscured populations \citep{R.Li_2025,Hu_2025}.

Similar to the luminosity ratio, the BH-to-stellar mass ratio can be evaluated analytically as 
\begin{align}
\mathscr{M}&\simeq 0.15~ \xi \zeta^{-1}
t_{\rm fb,1}^{1/3}
 T_{\rm vir,6}^{1/2}\left(\frac{1+z}{10}\right)^{17/6},\nonumber\\[3pt]
 & \propto (1+z)^{17/6-9\omega/16}\longrightarrow \left(\frac{1+z}{10}\right)^{1.15}.
 \label{eq:M_ratio}
\end{align}
The redshift dependence of the BH-to-stellar mass ratio is steeper than that of the luminosity ratio (see Figures~\ref{fig:L_ratio} and \ref{fig:M_ratio}),
but follows a similar trend.

Figure~\ref{fig:LRD_condition} illustrates the halo conditions that give rise to LRD phases in the $z$-$M_{\rm h}$ plane.
The solid curves indicate the optical-to-UV luminosity ratio at different halo conditions where a LRD emerges: $L_{\rm opt}/L_{\rm UV}=1$ (the LRD color boundary), 
$2$ (typical LRDs), $4$, and $10$ (extremely red LRDs). 
The gray curves trace the assembly histories of quasar-host galaxies (dashed) and less-massive galaxies (solid).
Along the halo assembly track, the optical-to-UV luminosity ratio remains nearly constant at high redshifts and decreases toward lower redshifts, as also shown in Figure~\ref{fig:L_ratio}.
The region filled with red dots marks the halo conditions where the BH envelope and nuclear starburst together produce the characteristic V-shaped SEDs of LRDs with 
$L_{\rm UV}\gtrsim 10^{10}~\lsun$ (corresponding to ${\rm SFR}_{\rm nuc}\gtrsim 1.0~\msunyr$);
\begin{align}
1  \lesssim  &~ L_{\rm opt}/L_{\rm UV}\lesssim 10, \nonumber \\[4pt]
2\times 10^{10} & \lesssim M_{\rm h}/\msun \lesssim 10^{12}.
\end{align}
Finally, the light-gray shaded region highlights the onset of feedback-free starburst modes \citep[e.g.,][]{Dekel_2023,Li_2024_FFB,Dekel_2025a}, where significantly redder SEDs dominated by the BH-envelope emerge though 
such cases are rarer.

\vspace{2mm}

\section{Discussion}\label{sec:discussion}

\subsection{Broad-line regions}\label{sec:BLR}

While the BH-envelope model successfully reproduces several key features of observed LRD spectra 
(e.g., red optical continua, absence of X-rays and radio jets, presence of Balmer break, and weak hot dust emission),
the origin of broad-line emission has not been addressed so far. 
The difficulty arises because an optically thick, low-temperature envelope that largely covers 
the central accreting BH would prevent ionizing photons from escaping.

One possible solution is to consider a non-spherical geometry, where ionizing radiation 
from the central accretion disk can leak through relatively low-density polar regions of the envelope \citep[e.g.,][]{Liu_2025b,X.Lin_2025_Egg}. 
Such a configuration could result from the angular momentum of inflowing gas. 
However, this geometrical effect would also imply a strong viewing-angle dependence 
in the observed LRD spectra. 
This is inconsistent with the observational fact that LRDs show a remarkably uniform 
V-shaped SED and a consistent lack of detectable X-ray emission across viewing directions.
Furthermore, since X-rays are more penetrating than UV photons, any radiation leakage that explains the observed UV continuum would inevitably lead to detectable X-ray emission, which is not observed.

In contrast, the composite SED model discussed in this paper offers an alternative explanation for 
the origin of broad-line emission through ionizing photons produced by young starbursts forming 
outside the gaseous envelope.
Assuming Case~B radiative recombination for atomic hydrogen, the H$\alpha$ luminosity can be related to 
the star formation rate as 
\begin{equation}
    L_{\rm H\alpha} =6.9\times 10^{41}~{\rm erg~s}^{-1}\left(\frac{\rm SFR_{\rm nuc}}{\msunyr}\right),
\end{equation}
where the conversion factor is based on a $\zsun/50$ stellar population 
(see Table~7 of \citealt{Schaerer_2002}).
A recent analysis for spectroscopically confirmed LRDs found that both broad and narrow H$\alpha$ luminosities are linearly correlated 
with the UV continuum with ratios consistent with that observed in star forming galaxies rather than in nearby unobscured AGNs \citep{Asada_2026}.
Using Equation~(\ref{eq:SFR}), the H$\alpha$ luminosity is given as
\begin{equation}
    L_{\rm H\alpha} =3.1\times 10^{42}~{\rm erg~s}^{-1}
    ~\zeta T_{\rm vir,6}^{3/2}\left(\frac{1+z}{10}\right)^{-3/2}.
\label{eq:LHa}
\end{equation}
This predicted H$\alpha$ luminosity is consistent with values measured for broad H$\alpha$ emitters
\citep[][without dust correction]{Harikane_2023_agn}, but is lower by a factor of $\simeq 5$ compared to 
brighter broad H$\alpha$ emitters identified by slitless spectroscopy \citep{Matthee_2024}.
We note that the estimate in Equation~(\ref{eq:LHa}) gives a lower limit because
LRDs can emerge in more massive halos at lower redshifts (i.e., $L_{\rm H\alpha} \propto M_{\rm h}$).
Since cold gas collapses to the halo center for $M_{\rm h}\lesssim 10^{12}~\msun$ \citep{Dekel_2006},
the H$\alpha$ luminosity can reach as high as 
$L_{\rm H\alpha} \simeq 3\times 10^{43}~{\rm erg~s}^{-1} ~\epsilon_{-1} \tau_{1}^{-1}f_{\rm d,0.3}$.
The most luminous LRD reported by \citet{Labbe_2024b} has $L_{\rm H\alpha}\simeq 10^{44}~{\rm erg~s}^{-1}$, 
which is achievable with higher disk mass fractions and star formation efficiencies,
such that $\epsilon_\star f_{\rm d}\gtrsim 0.1$.

The width of the H$\alpha$ emission line is determined by the virial velocity at, or just outside, the envelope,
as estimated in Equation~(\ref{eq:Vph}).
The relation between FWHM and virial velocity depends on the geometry of broad-line region clouds,
with $V_{\rm vir}^2 = f_{\rm vir}({\rm FWHM})^2$, where $f_{\rm vir}$ is the virial factor.
Assuming a spherical distribution of clouds ($f_{\rm vir}=4/3$; \citealt{Netzer_1990}), we obtain
\begin{equation}
    {\rm FWHM}\simeq 1,000~\kms~\lambda_{\rm Edd}^{-1/4} M_6^{1/4}T_0.
\end{equation}
This value matches the threshold typically used to define broad-line emission (${\rm FWHM}\geq 1,000~\kms$),
but does not reproduce significantly broader emission lines with ${\rm FWHM}\simeq 2,000-3,000~\kms$
unless the BH mass is as high as $\sim 10^8~\msun$.
However, such high BH masses are disfavored because AGN momentum and radiative feedback
would suppress the LRD phase (see Section~\ref{sec:AGNfeedback}).

Another plausible mechanism to broaden line widths is electron scattering.
Moderate scattering of line emission by free electrons with thermal velocity 
$v_{\rm th,e}=\sqrt{2k_{\rm B}T/m_{\rm e}}\simeq 550~\kms$ for $T=10^4~\K$, where $m_{\rm e}$ is the electron mass, 
should play a role in shaping line spectra, since a subset of LRDs with high signal-to-noise ratios indeed exhibit 
exponential line profiles, a clear signature of electron scattering.
As shown in \citet{Chang_2025_scattering} (see also \citealt{Rusakov_2025}), 
line widths of ${\rm FWHM}\simeq 2,000-3,000~\kms$ can be maintained through moderate electron scattering with 
optical depth of $\tau_{\rm es}\simeq 2-3$ for $T=10^4~\K$, even if the intrinsic line width is only $\simeq 1,000~\kms$.
The required optical depth corresponds to an electron column density of $N_{\rm e}\simeq {\rm a~few}\times 10^{24}~{\rm cm}^{-2}$,
or equivalently an electron density $n_{\rm e}\simeq 10^7~\cc (N_{\rm e}/3\times 10^{24}~{\rm cm}^{-2})(r/0.1~\pc)^{-1}$.
Such column densities can naturally arise in optically thick winds launched from the BH-envelope system.
Moreover, clouds of neutral hydrogen with high density of $n_{\rm H}\gtrsim 10^{9}-10^{11}~\cc$ are expected to form in LRDs
as indicated by the presence of prominent absorption features on top of broad Balmer lines 
\citep[e.g.,][]{Matthee_2024,Lin_2024,Juodzbalis_2024,Wang_2025,Kocevski_2025,DEugenio_2026} as well as extremely deep Balmer break features
\citep{Inayoshi_Maiolino_2025,Ji_2025,Naidu_2025,deGraaff_2025b,Taylor_2025b}.

Line broadening by electron scattering through nuclear dense plasma is a key mechanism in our composite SED model to account for broad-line AGNs with ${\rm FWHM}\gtrsim 2,000-3,000~\kms$.
Nevertheless, whether this mechanism operates effectively in real systems remains uncertain.
A more detailed treatment of line broadening within our framework is needed for future investigation.
In addition, detailed analyses of emission and absorption features will be essential to critically test theoretical models \citep[e.g.,][]{Brazzini_2025,Chang_2025_scattering}; 
in some cases, the dynamical BH mass measurement appears consistent with the single-epoch virial BH mass \citep{Juodzbalis_2025} but inconsistent with a model assuming electron scattering \citep{Rusakov_2025}.

Finally, we remark that a subset of LRDs identified by JWST/NIRCam photometry
shows a characteristic V-shaped SED but only narrow H$\alpha$ emission with
${\rm FWHM}\simeq 200-500~\kms$ \citep{Zhang_2025_NLR}.
Importantly, these objects share similar color and morphologies with broad-line LRDs,
differing mainly in their line widths.
Within the framework of coevolving BH-starburst systems, the narrow lines imply rapidly growing low-mass BHs of $M_{\rm BH}<10^5~\msun$.
Such phases occupy only $\sim 10\%$ of the LRD lifetime (see Figure~\ref{fig:BHgrowth}) compared to systems with BHs near the momentum-feedback critical mass, yet remain frequent enough to be observable.

\subsection{Lack of highly ionized elements}\label{sec:ionph}

To investigate the origin of the UV emission (and associated ionizing photons) observed in LRDs,
it is useful to examine emission lines with ionization potentials higher than the hydrogen ionization threshold
$E_{\rm H}=13.6~{\rm eV}$ (e.g., \ion{C}{4}~$\lambda1549$ and [\ion{Ne}{5}]~$\lambda3426$; \citealt{Lambrides_2024}).
Among the standard diagnostics employed in AGN searches, the \ion{He}{2} lines (\ion{He}{2}~$\lambda 1640$ and 
\ion{He}{2}~$\lambda 4686$) are particularly informative \citep[e.g.,][]{Nakajima_2022,Ubler_2023}.

Intriguingly, \ion{He}{2}~$\lambda 4686$ emission has been marginally detected in some LRDs
but with very low line ratios of \ion{He}{2}~$\lambda 4686$/H$\beta \sim 10^{-2}$,
i.e. less than 5\% of the median value for local AGN \citep{Wang_2025b}.
This observational result strongly constrains the ionizing spectral shape of LRDs.
Their spectra must include hydrogen-ionizing photons, but lack harder radiation needed to doubly ionize helium.
This picture is broadly consistent with the schematic structure of LRDs shown in Figure~\ref{fig:cartoon}.
In particular, young, massive stars with very low metallicity ($\simeq \zsun/50$ but nonzero) yield 
\ion{He}{2}~$\lambda 4686$/H$\beta \simeq 8\times 10^{-4}$ or a ratio between the hydrogen/ionized helium ionizing photon number fluxes,
$Q({\rm He}^+)/Q({\rm H})\simeq 5\times 10^{-4}$ \citep{Schaerer_2002}, consistent with the observed weakness 
of \ion{He}{2}~$\lambda 4686$ emission.

On the other hand, we remark that a few LRDs indeed show high-ionization lines. For instance, \citet{Tang_2025} reported narrow
\ion{N}{5}~$\lambda$1240 emission in an LRD at $z=6.98$, although neither \ion{C}{4} nor \ion{He}{2} emission lines were observed. 
Because \ion{N}{5} is unlikely produced by stellar processes, its detection likely requires nonstellar ionizing photons.
Possible explanations are that the AGN radiation escapes through a funnel-like geometry aligned with the rotation axis \citep[e.g.,][]{X.Lin_2025_Egg,Tang_2025},
or that these lines are produced from post-shocked, hot plasma in the LRD nuclei (S.~Takasao \& K.~Inayoshi 2026, in prep).

\subsection{X-ray emission}\label{sec:Xray}

X-rays are produced by SN blast waves when the shocked material cools radiatively.
During the cooling and shell formation phase ($t_{\rm sf}\sim 100-400~{\rm yr}$; see Equation~\ref{eq:tsf} for $Z\simeq \zsun/50$), 
multiple SNe with $N_{\rm SN}^{\rm sf}\sim3$ can occur given the expected SN rate. 
Since the post-shock temperature is $T_{\rm sf}\sim10^7$~K, most of the kinetic energy is radiated via X-ray emission, 
yielding a luminosity of
\begin{equation}
L_{\rm X} \sim N_{\rm SN}^{\rm sf} E_{\rm SN}/t_{\rm sf} \sim {\rm a~few}\times 10^{41}~{\rm erg~s}^{-1}.
\end{equation}
This X-ray luminosity lies below the upper limits of $L_{\rm X}\lesssim{10}^{43}-10^{44}~{\rm erg}~{\rm s}^{-1}$ derived from Chandra 
non-detections of individual LRDs, and is comparable to that from stacked LRDs \citep{Yue_2024,Maiolino_2025,Sacchi_2025}.
In LRDs with more active star formation and brighter UV emission, the SN rate would be higher, potentially enhancing the detectability 
of X-rays from SN blast waves.
X-ray binaries may also contribute, though their X-ray luminosity is expected to reach only
$L_{\rm X}\sim 4\times 10^{40}~{\rm erg~s}^{-1}({\rm SFR}/10~\msunyr)$ if the local empirical relation holds at high redshifts 
\citep[e.g.,][]{Grimm_2003,Mineo_2014}; the $L_{\rm X}/{\rm SFR}$ ratio could be higher in the early universe 
\citep[][]{Dijkstra_2012,Fragos_2013a}.
A more comprehensive evaluation of these contributions is beyond the scope of this work and will be left for future investigation.

\subsection{Time variability}\label{sec:Tvar}

Variability is a key diagnostic in AGN studies, as it directly traces emission from compact accreting BHs.
For LRDs, however, the detection of variability is challenging due to the lack of dedicated multi-epoch JWST observations, and 
most sources show no significant flux changes \citep[e.g.,][]{Kokubo_Harikane_2025,Zhang_2025}.
Nonetheless, a small fraction exhibit clear variability at the level of $\sim 0.3-0.8$ mag on timescales of $\lesssim 10-100$ days, providing strong evidence that at least some LRDs host compact, actively accreting BHs \citep{Zhang_2025,Ji_2025,Furtak_2025}.

In our two-component SED model, the optical variability originates from the convective envelope surrounding the BH.
Linear stability analyses suggest that such low-temperature envelopes excite opacity-driven pulsations, commonly referred to as the $\kappa$-mechanism 
\citep[e.g.,][]{Li_Gong_1994,Heger_1997,Baraffe_2001}.
For the envelopes of supermassive stars with $M_\star \gtrsim 10^{3}-10^{5}~\msun$ \citep{Inayoshi_2013a,Hosokawa_2013}, the fundamental pulsation mode is unstable against and has a characteristic period of
\begin{align}
    t_{\rm dyn}\equiv 2\pi \sqrt{\frac{R_{\rm ph}^3}{GM_{\rm BH}}} \simeq 38~{\rm yr}~\lambda_{\rm Edd}^{3/4}M_6^{1/4}T_0^{-3}.
\end{align}
This timescale is too long to be observed within a few years in the observer's frame, unless higher-order pressure-modes with shorter periods become unstable.

In systems composed of multiple objects, each exhibiting only small intrinsic variability, observable fluctuations are further suppressed due to de-synchronization.
Within our framework, however, UV variability could arise from relatively synchronized SNe occurring during the feedback timescale, $t_{\rm fb}\sim 10-20~{\rm Myr}$.
Given a SN rate of $\sim 0.01~{\rm yr}^{-1}$ in the source frame, monitoring several hundred high-redshift LRDs would statistically result in one LRD hosting a single SN event.
If such an event is a core-collapse SN with a typical peak luminosity of $\simeq 10^{43}~{\rm ergs}^{-1}$, comparable to the UV luminosity of LRDs, the UV continuum of the host LRD could vary by up to a factor of two.

%%%%%%%%%%
%%% Fig. 9 %%%
%%%%%%%%%%
\begin{figure}
\centering
\includegraphics[width=84mm]{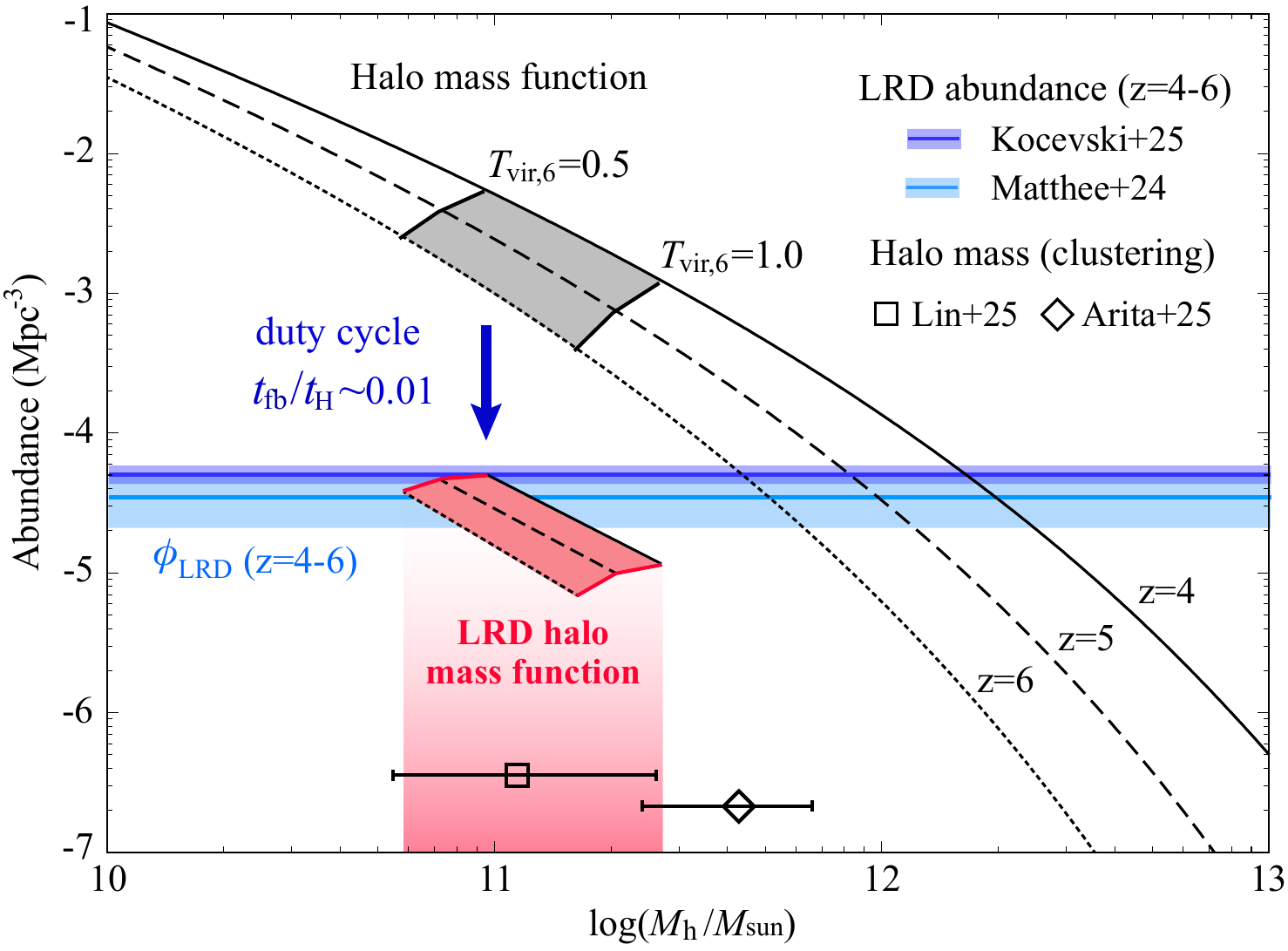}
\caption{Mass function of dark matter halos hosting LRDs at $z=4-6$, adopting virial temperatures of $T_{\rm vir,6}=0.5$ and $1.0$. 
The LRD abundance is given by multiplying the halo abundance by the duty cycle, defined by
$f_{\rm duty} = t_{\rm fb}/t_{\rm H}$.
The predicted values agree with the observed LRD abundances for photometrically selected samples \citep[blue,][]{Kocevski_2025} 
and spectroscopically confirmed samples with broad H$\alpha$ emission \citep[cyan,][]{Matthee_2024}.
These results suggest that LRD reside in halos with $M_{\rm h}\simeq 6\times 10^{10}-3\times 10^{11}~\msun$ at $z\simeq 4-6$,
which are comparable or lower than the halo masses inferred from clustering analyses of JWST-selected AGNs 
including LRDs and unobscured populations \citep{Arita_2025,X.Lin_2025_cluster}.
}
\label{fig:hmf}
\vspace{1mm}
\end{figure}

\subsection{Cosmic abundance and halo mass of LRDs}\label{sec:phi}

The cosmic abundance of observed LRDs rises from $z\gtrsim 10$, peaks at $z\sim 4-8$, and declines toward lower redshifts \citep{Kocevski_2025,Kokorev_2024a,Inayoshi_2025a,Ma_2025a,Zhuang_2025,Euclid_LRD_2025,Taylor_2025b,Tanaka_2025_z10LRD}.
Around the peak, the number density of LRDs remains nearly constant at $\phi_{\rm LRD} \simeq 6\times 10^{-5}~\mpc^{-3}$.
This low abundance corresponds to that of very massive dark matter halos with $M_{\rm h}\sim 10^{12}-10^{13}~\msun$, 
according to the Sheth-Tormen halo mass function at $z=4-6$ (Figure~\ref{fig:hmf}).
However, clustering analyses for LRDs suggest substantially lower host halo masses of $M_{\rm h}\simeq 5\times 10^{11}~\msun$ \citep{Arita_2025}, 
based on 27 low-luminosity broad-line AGNs, $\sim 20\%$ of which are LRDs.
Since LRDs are expected to reside in less massive halos than unobscured quasar hosts \citep{Pizzati_2025},
the true host halo masses are likely even lower.
Indeed, \citet{X.Lin_2025_cluster} estimate $M_{\rm h}\simeq 1.1\times 10^{11}~\msun$ from clustering of
23 low-luminosity broad-line AGNs with $\sim 50\%$ identified as LRDs.

In our model, dark matter halos with virial temperatures of $T_{\rm vir}=5\times 10^5- 10^6~\K$ are likely LRD hosts.
Using these values, the upper limit of the LRD abundance is given by assuming a duty cycle
defined as the ratio of the LRD lifetime to the Hubble time,
\begin{equation}
f_{\rm duty} \equiv \frac{t_{\rm fb}(T_{\rm vir})}{t_{\rm H}(z)} \simeq 0.013 \left(\frac{1+z}{6}\right)^{3/2},
\end{equation}
for $T_{\rm vir,6}\simeq 0.5-1$.
The predicted LRD abundance matches both photometrically selected samples \citep[blue,][]{Kocevski_2025} 
and spectroscopically confirmed samples with broad H$\alpha$ emission \citep[cyan,][]{Matthee_2024}.
This analysis predicts that typical LRD hosts are dark matter halos with $M_{\rm h}\simeq 6\times 10^{10}-3\times 10^{11}~\msun$ at $z\simeq 4-6$ (see the red thick curves in Figure~\ref{fig:z_growth}).
The predicted halo mass range agrees with that based on clustering analyses with a large LRD sample \citep{X.Lin_2025_cluster}.

Furthermore, if the UV continuum emission of LRDs originates from compact, young stellar clusters in the nuclear regions 
as proposed in this paper, the UV-luminosity function of LRDs reflects that of the host galaxies \citep{Asada_2026}.
The abundance of Lyman-break galaxies at similar UV magnitudes of $-20 \lesssim M_{\rm UV}/{\rm mag}\lesssim -18$ 
\citep[e.g.,][]{Harikane_2022a} is consistent with that of dark matter halos that satisfy the conditions for LRD emergence ($T_{\rm vir,6}\simeq 0.5-1$).

\subsection{Dust and metal production}\label{sec:metal}

Energetic SNe eject heavy elements into the surrounding material, and dust grains likely form in dense expanding shells.
Assuming a dust yield with $m_{\rm dust}\sim 0.1-1.0~\msun$ per core-collapse SNe, the total dust mass produced within the nuclear region is estimated as
\begin{equation}
    M_{\rm dust}\simeq 5\times 10^4~\msun ~\epsilon_{-1}t_1 f_{\rm d,0.3}f_{\rm n,0.1},
\end{equation}
where $m_{\rm dust}=0.3~\msun$ is adopted \citep{Schneider_Maiolino_2024}.
The duration before SN feedback clears gas from the nucleus may be as long as $t\sim t_{\rm fb}\simeq 20~{\rm Myr}$,
if the starburst proceeds with a 100\% duty cycle.
Under such conditions, the dust mass can grow up to $M_{\rm dust}\sim 10^5~\msun$,
although this estimate neglects dust destruction in SN ejecta by reverse shocks
\citep[e.g.,][]{Nozawa_2006,Schneider_Maiolino_2024}.
In any case, the estimated dust mass is consistent with the dust mass in LRDs inferred from the IR SED fitting and related analysis
\citep{Casey_2025,K.Chen_2025}, which suggests modest dust extinction ($A_V \ll 1$ mag).
Moreover, if half mass of heavy elements remains in the gas phase, i.e., a 50\% of dust depletion factor, the nuclear metallicity ends up in $\sim \zsun/300$.
This value represents an upper limit under the assumption of continuous star formation with a 100\% duty cycle,
but is broadly consistent with the metallicity adopted in our SED analysis ($Z=\zsun/50$).

At later evolutionary stages, dust grains become non-negligible and begin to affect both the observed SEDs and the gas dynamics in the nuclear region.
In particular, hot dust heated by the central AGN, presumably toward the end of the LRD phase, re-emits a remarkably red continuum extending from the rest-frame optical to the mid-infrared \citep[e.g.,][]{AlonsoHerrero_2006,RamosAlmeida_2011,Ichikawa_2014,Hickox_Alexander2018,Lyu_2022}.
This transition from dust-poor to dust-rich environments reflects the accumulation of dust grains through SN feedback, gradually giving a rise to a classical AGN dust torus \citep{Z.Li_2025}.
When a centrally concentrated dust torus is established, as seen in nearby AGNs, there is a dichotomy between type 1 and type 2 AGN populations owing to the geometric arrangement \cite[e.g.,][]{RamosAlmeida_2017,Hickox_Alexander2018}.

Finally, we point out that LRDs offer a unique laboratory to probe low-metallicity massive stellar populations via elemental abundance ratios in their AGN environments \citep[e.g.,][]{Maiolino_Mannucci_2019}.
In particular, the \ion{Fe}{2}/\ion{Mg}{2} flux ratio, widely used as a proxy for [Fe/Mg], is observed to be high and nearly redshift-independent in luminous AGNs up to $z\sim 7$ \citep[e.g.,][]{Nagao_2006b,Onoue_2020,Schindler_2020,Jiang_2024}.
This fact implies rapid iron enrichment possibly driven by massive-star explosions (e.g., Type-II SNe and hypernovae) rather than by Type-Ia SNe \citep{Umeda_2008,Nomoto_2013,Toyouchi_2022}.
However, current measurements of \ion{Fe}{2}/\ion{Mg}{2} ratios at high redshifts remain poorly constrained due to non-detections of these UV emission lines.
An exception is the brightest LRD, UNCOVER-45924 \citep[][]{Labbe_2024b}, which exhibits prominent \ion{Fe}{2} and \ion{Mg}{2} emission lines with a flux ratio of \ion{Fe}{2}/\ion{Mg}{2}$\sim 10$, consistent with values seen in luminous quasars. 
Future JWST UV spectroscopy of LRDs offers a unique opportunity to directly trace the onset of chemical enrichment (see also \citealt{Trefoloni_2025}), constrain the pathway from LRDs to normal AGNs, and place stringent limits on early stellar evolution and the initial mass function.

\section{Summary}\label{sec:summary}

JWST deep surveys have recently revealed a new population of AGNs known as 
little red dots (LRDs) \citep[e.g.,][]{Matthee_2024,Greene_2024,Labbe_2025,Kocevski_2025}.
They show broad Balmer lines and distinctive V-shaped SEDs, while classical AGN signatures such as hard X-rays, hot dust emission, and radio jets are weak or absent.
These properties likely reflect dense gaseous environments surrounding the central BHs \citep{Matthee_2024,Maiolino_2024_JADES,Lin_2024,Inayoshi_Maiolino_2025,Ji_2025}. 
In particular, if the accreting BH is enshrouded by an optically thick gas envelope 
with a surface temperature of $T_{\rm eff}\simeq 5000~\K$, characteristic of the Hayashi line \citep{Hayashi_1961}, 
the red optical continuum and several other spectral features can be reproduced in a self-consistent way.
However, such an envelope alone produces little UV continuum and broad-line emission 
due to its high covering factor \citep{Kido_2025,Liu_2025b,X.Lin_2025_Egg}.
Although young stellar populations have been proposed as an additional UV source, 
simple two-component models still face difficulty in explaining the uniform V-shaped SEDs of LRDs \citep[e.g,][]{Wang_2024b,Setton_2025a,Kocevski_2025}.

In this paper, we investigated the properties of mass accretion in early protogalaxies that lead to both a BH-envelope system and concurrent nuclear starbursts, 
and proposed a two-component model for the V-shaped SEDs of LRD populations (see a schematic view in Figure~\ref{fig:cartoon}).
In this model, the red optical continuum arises from the Wien tail of the blackbody spectrum emitted by the envelope surface, without significant dust reddening. 
The observed rest-frame optical luminosity ($L_{\rm opt}\simeq 10^{44}~{\rm erg~s}^{-1}$) can be reproduced by the Eddington luminosity of a BH with $M_{\rm BH}\simeq 10^6~\msun$ for $A_V=0$.
The observed rest-frame UV continuum emission ($L_{\rm UV}\simeq 10^{43}~{\rm erg~s}^{-1}$) requires intense starburst activity with ${\rm SFR}\gtrsim 1~\msunyr$,
leading to the formation of a $\sim 10^7~\msun$ stellar cluster within $\sim 10$ Myr at $\lesssim 100~\pc$.
These two components naturally explains the V-shaped SEDs of both the stacked LRD spectrum 
and JWST/NIRSpec PRISM spectra of individual sources 
\citep[][]{Furtak_2024,Akins_2025,Ji_2025,Naidu_2025,deGraaff_2025b,Labbe_2024b,Taylor_2025b}.

Massive stars formed in such active bursts undergo SN explosions and inject energy and momentum into the surrounding medium.
This feedback process expels gas from the nucleus, quenches accretion onto the BH, and thus regulates the LRD activity,
leading to a transition into normal AGN phases in the subsequent accretion stages \citep{Inayoshi_2025a}.
Super-Eddington mass growth of the BHs through the LRD envelope structure last for $\simeq 15~{\rm Myr}$, one-third of the Salpeter timescale,
until quenching by stellar feedback.
By the time when the LRD phase terminates, the BH mass reaches a critical mass of $\sim 10^7~\msun$ (Equation~\ref{eq:critLRD}).
We pointed out that this should serve as a theoretical upper limit, which aligns with the inferred cutoff luminosity of the LRD luminosity function \citep{Ma_2025b}.

The stellar/AGN feedback prescription is incorporated into the framework of high-$z$ halo assembly 
(see Figures~\ref{fig:z_growth} and \ref{fig:BHgrowth}).
Our model predicts that the optical-to-UV luminosity ratio of the BH+starburst components rises rapidly
and saturates at $L_{\rm opt}/L_{\rm UV}\sim 2-10$, depending on the parent dark matter halo mass.
This luminosity ratio is broadly consistent with that observed for the LRD population (Figure~\ref{fig:L_ratio}) and remains nearly constant over $4<z<10$.
At lower redshifts of $z<4$, however, the luminosity ratio decreases, causing LRDs to lose their distinctive colors 
and evolve into galaxy-light-dominated systems.
This framework naturally explains the origin of the uniform SEDs, their redshift evolution including disappearance, 
and their connection to parent halo properties.

Similar to the luminosity ratio, the ratio of BH mass to nuclear stellar mass at the end of LRD phase also
converges to a narrow range at each redshift, with $M_{\rm BH}/M_{\star,\rm nuc}\simeq 0.1$ for $5<z<10$.
Toward lower redshifts ($z<1$), the ratio shifts increasingly toward stellar dominance and eventually approaches the local value of $\sim 0.1\%$ \citep[e.g.,][]{Kormendy_Ho_2013,Reines_Volonteri_2015,Greene_ARAA_2020}.

This framework also accounts for the origin of broad-line emission in LRDs.
Massive young stars in the nuclear regions (outside the BH envelope) produce sufficient ionizing radiation to power H$\alpha$ luminosities of $L_{\rm H\alpha}\simeq 10^{42}-10^{43}~{\rm erg~s}^{-1}$, consistent with observations.
The intrinsic line widths are expected to be up to ${\rm FWHM}\simeq 1,000~\kms$, limited by the circular 
velocity outside the envelope.
To reproduce the broader lines of $\gtrsim 2,000-3,000~\kms$, modest electron scattering
(or additional transfer effects such as Balmer-line resonance scattering; see e.g.,
\citealt{Rusakov_2025} and \citealt{Chang_2025_scattering}) would be required.

In summary, LRDs can be viewed as a transient phase of super-Eddington accreting BHs enshrouded by red gaseous envelopes
inflated to the Hayashi limit, and persisting for only one-third of a Salpeter time before feedback quenches their activity --- perhaps 
the first luminous red glow of infant BHs.

\begin{acknowledgments}
We greatly thank Avishai Dekel, Seiji Fujimoto, Jingsong Guo, Yuichi Harikane, Luis Ho, Kunihito Ioka, Linhua Jiang, 
Tatsuya Matsumoto, Masaru Shibata, and Takumi Tanaka for constructive discussions.
K.I. acknowledges support from the National Natural Science Foundation of China (12573015, W2532003), the Beijing Natural Science Foundation (IS25003), 
and the China Manned Space Program (CMS-CSST-2025-A09).
This work was supported by JSPS KAKENHI grant Nos. JP24K00668, JP23H04899, JP22H00130 (K.K.).
We thank the Yukawa Institute for Theoretical Physics at Kyoto University. 
Discussions during the YITP workshop YITP-W-25-08 on Exploring Extreme Transients were useful to complete this work.
K.I. thanks the study room at the Wanliu campus of Peking University, which provided a quiet and comfortable environment for completing this work during the summer.

This work is based on observations made with
the NASA/ESA/CSA James Webb Space Telescope.
The data were obtained from the Mikulski Archive for
Space Telescopes at the Space Telescope Science Institute, which is operated by the Association of Universities
for Research in Astronomy, Inc., under NASA contract
NAS 5-03127 for JWST.

\end{acknowledgments}

%\begin{contribution}
%All authors contributed equally to the Terra Mater collaboration.
%\end{contribution}
%\facilities{HST(STIS), Swift(XRT and UVOT), AAVSO, CTIO:1.3m, CTIO:1.5m, CXO}

\newpage

\appendix

\section{Supernova bubble evolution}

A young, massive stellar population formed within a compact disk owing to gravitational instability
produces blue UV continuum emission with brightness associated with the star formation rate, stellar age, and metallicity.
In our fiducial case, where ${\rm SFR}\sim 1.0~\msunyr$, $t_{\rm age}=10~{\rm Myr}$, and $Z=\zsun/50$,
the UV luminosity of $L_{\rm UV}\simeq 10^{10}~\lsun$ becomes comparable to the typical UV brightness observed in LRDs.

As a natural consequence, massive stars in the nuclear star forming region will undergo SN explosions (SNe),
which inject mass, energy, and momentum into the surrounding medium, and eventually would self-regulate star formation activity itself
and gas feeding to the central BH.
Following the analytic theory summarized in \citet{Kim_Ostriker_2015}, we discuss the dynamics of radiative supernova remnants (SNRs) 
in the simplest case of spherical expansion under a uniform ambient medium with hydrogen number density of $n_{\rm amb}$ \citep[see][]{McKee_Ostriker_1977}.
In analytic models, each evolutionary stage of the SNR expansion can be approximated by a power-law form of the SNR size in time
as $r_{\rm SNR}\propto t^{\delta}$. 
The value of the power-law index $\delta$ defines four main stages of the expansion \citep{Cioffi_1988, Ostriker_McKee_1988,Draine_2011}:
free expansion ($\delta =1$), the Sedov-Taylor phase ($\delta =2/5$), pressure-driven snowplow ($\delta =2/7$), and momentum-conserving
snowplow ($\delta =1/4$). Note that the power-law index decreases at each successive stage, where the SNR expansion decelerates.

Let us consider a SN explosion ejecting a mass of $M_{\rm ej}$ with a kinetic energy of $E_{\rm SN}=10^{51}E_{51}~{\rm erg}$.
In the initial stage, the ejecta expands nearly freely with a constant velocity, $v_{\rm ej}=(2E_{\rm SN}/M_{\rm ej})^{1/2}$,
so that the radius of the SNR increases with time, $r_{\rm SNR}\propto r$ ($\delta =1$).
During this free-expansion stage, the forward shock sweeps up the surrounding ambient medium, while the reverse shock 
propagates inward and begins to heat the ejecta.
The free-expansion phase ends when the swept-mass, $M_{\rm sw}=(4\pi/3)\rho_{\rm amb}r_{\rm SNR}^3$, becomes 
comparable to the initial ejecta mass, where $\rho_{\rm amb}=1.4~m_{\rm H}n_{\rm amb}$ for 10\% helium abundance. 
The deceleration radius of the SNR is 
\begin{equation}
    r_{\rm dec} = \left(\frac{3M_{\rm ej}}{4\pi \rho_{\rm amb}}\right)^{1/3}\simeq 0.088~\pc~ M_{\rm ej,1}^{1/3}n_5^{-1/3},
\end{equation}
where $M_{\rm ej,1}=M_{\rm ej}/(10~\msun)$ and $n_5 = n_{\rm amb}/(10^5~\cc)$, and the deceleration time is 
\begin{equation}
    t_{\rm dec} = \frac{r_{\rm dec}}{v_{\rm ej}}\simeq 27~{\rm yr}~ M_{\rm ej,1}^{5/6}E_{51}^{-1/2}n_5^{-1/3}.
\end{equation}

After $t>t_{\rm dec}$, when all of the ejecta are shocked, the SNR evolution can be approximated by a point-source explosion 
with an energy of $E_{\rm SN}$ embedded in medium with density of $n_{\rm amb}$.
Here, we consider the adiabatic phase, where radiative energy losses are irrelevant, neglecting the pressure in the ambient medium.
In this idealized setup, the SNR evolution is described by a self-similar solution scaled only a characteristic radius scale of 
$(E_{\rm SN} t^2/\rho_{\rm amb})^{1/5}$.
Imposing the jump conditions of the physical quantities across the shock front, a detailed solution is given by 
$r_{\rm SNR}=1.15167(E_{\rm SN} t^2/\rho_{\rm amb})^{1/5}$
for adiabatic gas of $\gamma=5/3$. Then, we obtain the shock radius, velocity, and post-shock temperature in the so-called 
Sedov-Taylor (ST) phase as
\begin{equation}
    r_{\rm ST} \simeq 0.20~\pc~E_{51}^{1/5}n_5^{-1/5}t_2^{2/5},
\end{equation}
\begin{equation}
    v_{\rm ST} \simeq 780~\kms ~E_{51}^{1/5}n_5^{-1/5}t_2^{-3/5},
\end{equation}
\begin{equation}
    T_{\rm ST} = 8.2\times 10^6~\K ~E_{51}^{2/5}n_5^{-2/5}t_2^{-6/5},
\end{equation}
where $t_2 = t/(100~{\rm yr})$ \citep{Draine_2011}.

The hot gas with $T\sim 10^7~\K$ radiates and looses its thermal energy. Due to the shock compression of the density, radiative cooling becomes important just behind the shock front and leads to the formation of a cold, dense shell.
Following \citet{Draine_2011}, we approximated the cooling function as $\Lambda = C T_6^{-\alpha}n_{\rm H}n_{\rm e}$
with $C\simeq 1.1\times 10^{-22}~{\rm erg~cm^3~s^{-1}}$ and $\alpha=0.7$ that gives a good fit for gas with $Z=\zsun$.
Using the cooling rate and estimating the time by when one third of the original explosion energy is lost radiatively,
the shell-formation time, radius of the SNR, shock velocity, and post-shock temperature are calculated as 
\begin{equation}
    t_{\rm sf}\simeq 88~{\rm yr}~E_{51}^{0.22}n_5^{-0.55},
    \label{eq:tsf}
\end{equation}
\begin{equation}
    r_{\rm sf}\simeq 0.19~{\rm pc}~E_{51}^{0.29}n_5^{-0.42}, 
\end{equation}
\begin{equation}
    v_{\rm sf}\simeq 900~\kms~E_{51}^{0.07}n_5^{0.13}, 
\end{equation}
\begin{equation}
T_{\rm sf}\simeq 1.1\times 10^7~\K~E_{51}^{0.13}n_5^{0.26}.
\end{equation}
Here, we note that the estimate above is based on the cooling function for $Z=\zsun$ gas, while 
our fiducial cases consider $Z\approx \zsun/50$.
With low-metallicity gas, where collision-induced emission of heavy elements is inefficient, 
the shell-forming time is delayed and the SNR size on the transition time is larger.
The metallicity dependence has been discussed and characterized with a power-law form as
$t_{\rm sf}\propto Z^{-5/14}$, $r_{\rm sf}\propto Z^{1/7}$, $v_{\rm sf}\propto Z^{3/14}$, and $T_{\rm sf}\propto Z^{3/7}$ 
for $T_{\rm sf}\lesssim 10^7~\K$ \citep{Cioffi_1988}. 
This metallicity dependence suggests $t_{\rm sf}\sim 100-400~{\rm yr}$ for our fiducial case ($Z=\zsun/50$).

Based on the analytical estimate, the momentum injection to the ISM from a single SN can be calculated.
In the ST stage, the strong blast wave heats and accelerates the ambient gas. Using the ST self-similar solution,
the radial momentum of the shocked gas is obtained as
\begin{equation}
    p_{\rm ST}\simeq 5.6\times 10^4~\msun \cdot \kms E_{51}^{4/5}n_5^{1/5}t_2^{3/5}.  
\end{equation}
The momentum at the shell-forming time is given as 
\begin{equation}
    p_{\rm sf}\simeq 1.1\times 10^5~\msun \cdot \kms E_{51}^{0.93}n_5^{-0.13}\left(\frac{Z}{\zsun/50}\right)^{-3/14}.
\end{equation}
For low-metallicity cases, since the shell-formation is delayed due to inefficient cooling, the blast wave in the ST accelerates the ejecta longer.

After the formation of a dense shell via radiative energy losses, the shell is still pushed outward by overpressured hot gas in the interior of the SNR. 
If the radiative cooling of the hot interior gas is still insignificant (i.e., the pressure of the interior hot gas is large enough), the SNR size is characterized by $r_{\rm SNR}\propto t^{2/7}$ and the momentum gradually increases with time as $t^{1/7}$. However, in reality, the hot interior begins to cool and it will change to $r_{\rm SNR}\propto t^{1/4}$. The resulting momentum injection saturates at 
\begin{equation}
    p_{\rm SNR}\simeq 1.62~p_{\rm sf}\simeq 1.8\times 10^5~\msun \cdot \kms E_{51}^{0.93}n_5^{-0.13}\left(\frac{Z}{\zsun/50}\right)^{-3/14},
    \label{eq:KO15}
\end{equation}
\citep{Kim_Ostriker_2015}.

\bibliography{ms}{}
\bibliographystyle{aasjournalv7}



\end{document}